\documentclass[aps,prx,showpacs,  amsmath,amssymb,longbibliography,showkeys,12pt]{revtex4-2}%
\usepackage{amsfonts}
\usepackage{amsmath}
\usepackage{amssymb}
\usepackage{graphicx}
\usepackage{epsfig}
\usepackage{exscale}
\usepackage{float}
\usepackage{bm}
\usepackage{suffix}
\usepackage{mathtools}
\usepackage{newunicodechar}
\usepackage{lipsum}
\usepackage{enumitem}
\usepackage{bbding}
\usepackage{tikz}
\usepackage{multirow}%
\usepackage[colorlinks=true,linkcolor=blue]{hyperref}
\usepackage{ifpdf }
\expandafter\ifx\csname package@font\endcsname\relax\else
 \expandafter\expandafter
 \expandafter\usepackage
 \expandafter\expandafter
 \expandafter{\csname package@font\endcsname}
\fi
\providecommand{\U}[1]{\protect\rule{.1in}{.1in}}

\DeclarePairedDelimiterX\MeijerM[3]{\lparen}{\rparen}
{\begin{smallmatrix}#1 \\ #2\end{smallmatrix}\delimsize\vert\,#3}
\newcommand\MeijerG[8][]{  G^{\,#2,#3}_{#4,#5}\MeijerM[#1]{#6}{#7}{#8}}
\WithSuffix
\newcommand\MeijerG*
[7]{
G^{\,#1,#2}_{#3,#4}\MeijerM*{#5}{#6}{#7}}

\begin{document}
\title[ ]{ High-Order Parametrization of the Hypergeometric-Meijer Approximants }
\author{Abouzeid M. Shalaby}
\email{amshalab@qu.edu.qa}
\affiliation{Physics Program, Department of Mathematics, Statistics and Physics, College of Arts and Sciences, Qatar University, P.O box 2713, Doha, Qatar}

\begin{abstract}
 In previous articles, we showed that, based on large-order asymptotic behavior,  one can approximate a divergent  series via the parametrization of a specific  hypergeometric approximant. The  analytical continuation is then carried out through a Mellin-Barnes integral representation of the hypergeometric approximant or equivalently using an equivalent form of the  Meijer G-Function. The parametrization process involves the solution of a non-linear set of coupled equations which is hard to achieve (might be impossible) for high orders using normal PCs.
In this work,  we  extend the approximation algorithm to accommodate any order (high or low)  of the given series in a short time. The extension also allows us to employ non-perturbative information like strong-coupling and large-order asymptotic data which are always used to accelerate the convergence. We applied the algorithm for different orders (up to O($29$))  of the ground state energy of the $x^4$ anharmonic oscillator with and without the non-perturbative information. We also considered   the available $20$ orders for the ground sate energy of the  $\mathcal{PT}-$symmetric $ix^{3}$ anharmonic oscillator as well as the given $20$ orders of its strong-coupling expansion or equivalently the Yang-Lee model. For high order weak-coupling parametrization, accurate results have been obtained for the ground state energy and the non-perturbative parameters describing strong-coupling and large-order asymptotic behaviors. The employment of the non-perturbative data accelerated the convergence very clearly.  The High temperature expansion for the susceptibility within the $SQ$ lattice has been also considered and led to  accurate prediction for the critical exponent and critical temperature.  
            
\end{abstract}
\maketitle

\section{ Introduction}

Frequently in physics, one is  confronted by the existence of divergent perturbation series. This can exist  in more than one type of series behavior. There exists divergent series with zero radius of convergence where perturbation fails to give reliable results for the whole complex plane of the perturbation parameter. Another type is a  series with finite radius of convergence but the region of interest is outside the   disk of convergence like critical region of high temperature expansion. Examples in physics for the first type   include (but not limited to)
the expansion of physical quantities within the $x^{4}$ anharmonic oscillator,
Ising model, the $\phi^{4}$ scalar field theory and QED. To draw reliable
results from such series one can apply resummation techniques like Borel
resummation \cite{Kleinert-Borel,zin-exp,Borel-6L,Borelg}, Pad$\acute{e}$ approximants
\cite{pade1, pade2,pade3} as well as variational methods \cite{Kleinert-Borel}.
Recently, a hypergeometric-Borel technique \cite{hyper-borel} has also been
applied to resum divergent perturbation series . The hypergeometric
approximants $_{\text{ }p}F_{p-1}$ \cite{prl-hyp,abohyp} have been also shown
to produce good approximations for a divergent series with zero-radius of
convergence. However, the approximants $_{\text{ }p}F_{p-1}$ have a series
expansion with finite-radius of convergence and thus when used to approximate
divergent series with zero-radius of convergence they show some shortcomings
\cite{hyper-borel,hyp1,hyp2}. In Refs.\cite{abo-hyyp-meij, abo-precize,eps7},
we introduced what we call it the hypergeometric-Meijer approximation algorithm. This algorithm can approximate different types of series based on the divergence manifestation. In fact, the type of divergence is manifested in  the growth factor of the series coefficients at large orders.  A series with finite radius of convergence has $0!$ growth factor while the zero radius of convergence ones can have $n!, (2n)!, \dots$ growth factors. Our   algorithm can treat such types of series and in fact it has the same spirit of the hypergeometric approximant introduced
\ by Mera et.al in Ref. \cite{prl-hyp} but in a way that respects the analytic
properties of the given-series and is able to accommodate all known
non-perurbative data associated with the given series. The employment of the
non-perturbative data is known to accelerate the convergence of resummation
techniques \cite{Kleinert-Borel} and it has been shown in our previous work that it is accelerating the convergence of the hypergeometric approximants as well. Our algorithm has been shown to give
excellent results for the approximation of different divergent perturbation series \cite{abo-hyyp-meij, abo-precize,eps7}.

The hypergeometric-Meijer algorithm is pretty simple (but accurate) and  has two main steps:
\begin{enumerate}
\item Approximating the given series with a hypergeometric series $_{\text{
}p}F_{q}$ that can be parametrized to reproduce all the known information about
the original series.
\item The parametrized  hypergeometric series is then analytically continued
using its integral representation in the form of a Mellin-Barnes integral or
equivalently in terms of a Meijer G function where \cite{HTF}:%
\begin{equation}
_{\text{ }p}F_{q}(a_{1},...a_{p};b_{1}....b_{q};z)=\frac{\prod_{k=1}^{q}%
\Gamma\left(  b_{k}\right)  }{\prod_{k=1}^{p}\Gamma\left(  a_{k}\right)  }%
\MeijerG*{1}{p}{p}{q+1}{1-a_{1},\dots,1-a_{p}}{0,1-b_{1},\dots,1-b_{q}}{z},
\label{hyp-G-C}%
\end{equation}
and
\begin{equation}
\MeijerG*{m}{n}{p}{q}{c_{1},\dots,c_{p}}{d_{1},\dots,d_{q}}{z}=\frac{1}{2\pi i}%
\int_{C}\frac{\prod_{k=1}^{n}\Gamma\left(  s-c_{k}+1\right)  \prod_{k=1}%
^{m}\Gamma\left(  d_{k}-s\right)  }{\prod_{k=n+1}^{p}\Gamma\left(
-s+c_{k}\right)  \prod_{k=m+1}^{q}\Gamma\left(  s-d_{k}+1\right)  }z^{s}ds.
\label{hyp-G-C2}%
\end{equation}
\end{enumerate}
For the hypergeometric approximants of interest $_{\text{ }p}F_{q}(a_{1},...a_{p};b_{1}....b_{q};z)$, where $p=q+1$ and $p=q+2$,  the above integral representation is known to converge \cite{HTF,abo-hyyp-meij}. 
To show how one can choose the suitable hypergeometric approximant for a given series, assume we
are given a series up to some order $n$ for a physical quantity $Q\left(
z\right)  $ in the form:%
\begin{equation}
Q\left(  z\right)  \approx\sum_{0}^{n}c_{i}z^{i}. \nonumber\label{petQM}%
\end{equation}
Based on its large-order behavior, a hypergeometric function $_{\text{ }%
p}F_{q}$, with a constraint on $L=p-q$, can be parametrized to accommodate all
known information given for the series under consideration. For a given series, one might know the first $n$ terms, the large-order asymptotic
behavior of the series and its asymptotic strong-coupling behavior. It is the
large-order behavior that determines the constraint on $L$. For instance, if
the given series has a finite radius of convergence then for large $i$,
$c_{i}$ behaves as $\sigma^{i}i^{b}$. The radius of convergence $R$ is then
$\frac{1}{\sigma}$. In this case, the suitable hypergeometric approximants are
$_{\text{ }p}F_{q}$ with $L=1$. In case the series has a zero-radius of
convergence with an asymptotic large-order behavior of the form $i!\sigma
^{i}i^{b}$, then the suitable approximants are $_{\text{ }p}F_{q}$, with $L=2$
and so on \cite{universal,abo-hyyp-meij, abo-precize,eps7}.

The hypergeometric approximant $_{p}F_{q}(a_{1},a_{2,}........a_{p};b_{1}%
,b_{2},....b_{q};\sigma z)$ has the series expansion:
\begin{equation}
_{\text{ }p}F_{q}\left(  {a_{1},a_{2,}....\ a_{p};b_{1},b_{2}....\ b_{q,}%
;\sigma x}\right)  =\sum_{n=0}^{\infty}\frac{\frac{\Gamma\left(
a_{1}+n\right)  }{\Gamma\left(  a_{1}\right)  }\frac{\Gamma\left(
a_{2}+n\right)  }{\Gamma\left(  a_{2}\right)  }.............\frac
{\Gamma\left(  a_{p}+n\right)  }{\Gamma\left(  a_{2}\right)  }}{n!\frac
{\Gamma\left(  b_{1}+n\right)  }{\Gamma\left(  b_{1}\right)  }\ \frac
{\Gamma\left(  b_{2}+n\right)  }{\Gamma\left(  b_{2}\right)  }.......\frac
{\Gamma\left(  b_{q}+n\right)  }{\Gamma\left(  b_{q}\right)  }\ \ }\left(
\sigma x\right)  ^{n}.
\end{equation}
For $L=1,2$ , we have shown that it can be parametrized to produce the  
asymptotic large order behavior \cite{universal,abo-hyyp-meij,
abo-precize,eps7} such that:
\begin{equation}
\sum_{i=1}^{p}a_{i}-\sum_{i=1}^{q}b_{i}-L=b.
\end{equation}
Also, the numerator parameters ($-a_{i}$ ) are representing the strong coupling
parameters of the given series \cite{abo-hyyp-meij}. However, technical
problems in the calculation arise for finite values of $z$ when the difference
$a_{k}-a_{j}$ is an integer \cite{Analytic2016}. Accordingly, as we will
explain later  when we impose the strong-coupling parameters into the
approximants to accelerate the convergence, it is more safer to employ the
first $a_{i}$\textquotesingle s with the difference $a_{k}-a_{j}$ is not an integer to
avoid singularities in the calculations.

Let us now show how to use the hypergeometric approximants to approximate a
given series. For simplicity, assume first that we have only the first five
orders of the perturbation series:
\[
Q\left(  z\right)  \approx\sum_{0}^{5}c_{i}z^{i},
\]
The weak-coupling parametrization assumes that we know the values of $c_{0},c_{1},c_{2},c_{3},c_{4}\ $\ and
$c_{5}$ but the non-perturbative parameters are not known. The ratio test can tel us about the radius of convergence of the
given series. If the given series is known to have a zero radius of
convergence with coefficients $c_i$ behave like $i!\sigma^{i}i^{b}$ for large $i$, then the suitable approximant is
\[
Q\left(  z\right)  \approx c_{0\text{ \ }3}F_{1}(a_{1},a_{2,}a_{3,}%
;b_{1};\sigma z).
\]
The approximant $_{\text{ \ }3}F_{1}(a_{1},a_{2,}a_{3,};b_{1};\sigma z)$ has
five parameters, namely $a_{1},a_{2,}a_{3,}b_{1}$and $\sigma$ to be determined.
Matching coefficients of same order of $z$ in the given series and the series
expansion of $c_0\  {}_3F_{1}$ we get;%

\begin{align}
c_{0\text{ \ }}\frac{a_{1}a_{2}a_{3}}{b_{1}}\sigma &  =c_{1}\nonumber\\
c_{0\text{ \ }}\frac{a_{1}a_{2}a_{3}\left(  a_{1}+1\right)  \left(
a_{2}+1\right)  \left(  a_{3}+1\right)  }{2!b_{1}\left(  b_{1}+1\right)
}\sigma^{2}  &  =c_{2}\nonumber\\
c_{0\text{ \ }}\frac{a_{1}a_{2}a_{3}\left(  a_{1}+1\right)  \left(
a_{2}+1\right)  \left(  a_{3}+1\right)  \left(  a_{1}+2\right)  \left(
a_{2}+2\right)  \left(  a_{3}+2\right)  }{3!b_{1}\left(  b_{1}+1\right)
\left(  b_{1}+2\right)  }\sigma^{3}  &  =c_{3}\nonumber\\
c_{0\text{ \ }}\frac{a_{1}a_{2}a_{3}\left(  a_{1}+1\right)  \left(
a_{2}+1\right)  \left(  a_{3}+1\right)  ......\left(  a_{1}+3\right)  \left(
a_{2}+3\right)  \left(  a_{3}+3\right)  }{4!b_{1}\left(  b_{1}+1\right)
....\left(  b_{1}+3\right)  }\sigma^{3}  &  =c_{4}\label{non-linear}\\
c_{0\text{ \ }}\frac{a_{1}a_{2}a_{3}\left(  a_{1}+1\right)  \left(
a_{2}+1\right)  \left(  a_{3}+1\right)  ......\left(  a_{1}+4\right)  \left(
a_{2}+4\right)  \left(  a_{3}+4\right)  }{5!b_{1}\left(  b_{1}+1\right)
....\left(  b_{1}+4\right)  }\sigma^{3}  &  =c_{5}\nonumber
\end{align}
This a set of non-linear algebraic equations can be solved for the five unknowns
$a_{1},a_{2,}a_{3},b_{1},\sigma$. The degree of non-linearity can be lowered
by generating the ratio $R_{n}=\frac{c_{n}}{c_{n-1}}$ and match it by the
corresponding ratio $g_{n}$ from the series expansion of the hypergeometric
approximant where
\[
_{p}F_{q}(a_{1},a_{2,}........a_{p};b_{1},b_{2},....b_{q};\sigma z)=\sum
_{n=0}^{\infty}h_{n}z^{n},
\]
and
\begin{equation}
g_{n}=\frac{h_{n}}{h_{n-1}}=\ \frac{%
{\displaystyle\prod_{i=1}^{p}}
\left(  a_{i}+n-1\right)  }{n%
{\displaystyle\prod_{j=1}^{q}}
\left(  b_{j}+n-1\right)  }\sigma. \label{hnn}%
\end{equation}
The set of equations $R_{n}=g_{n}$ is still non-linear and in going to higher
orders will make it very hard and might be impossible to solve it in a practical
time using normal computers. In fact, this represents a major obstacle that
prevents the current versions of \ hypergeometric-Meijer algorithm from
tackling the approximation of a divergent perturbation series with relatively high orders used as input. In literature,  one can find perturbation series obtained up to a relatively high order like the high-temperature expansion of Ising like models
\cite{HT1,HT2} for which the the hypergeometric approximants $_{p}%
F_{p-1}(a_{1},a_{2,}........a_{p};b_{1},b_{2},....b_{p-1};z)$ offer a good
approximation for the given series. Likewise, the ground state energy for both
hermitian $x^{4}$ \cite{x4-bender-69} and the non-Hermitian $ix^{3}$
\cite{ix3-bender-large} anhrmoinic oscillators are known up to high orders and
thus the parametrization of the hypergeometric approximants that can accommodate
information from the known orders is necessary. In this work, we introduce a
simple algorithm to get an equivalent (order by order) set of linear equations that can be
solved easily using a normal PC and for short time. Note that in Ref.
\cite{hyper-borel}, Mera et. al used the hypergeometric approximants
$_{p}F_{p-1}(a_{1},a_{2,}........a_{p};b_{1},b_{2},....b_{p-1};z)$ to
approximate the Borel series obtained by Borel transforming the given
perturbation series. They introduced the ansatz (Eq.(5) in the same reference)
to approximate the ratio $g_{n}$ for $_{p}F_{p-1}$. Although the important
idea of getting a linear set of equations followed by Mera et.al  is similar to
what we will follow in our Hypergeometric-Meijer algorithm, in our work
however, we do not use any ansatz and shall try to get a linear set of
equations not only for $_{p}F_{p-1}$ but for any hypergeometric approximant
$_{p}F_{q}$. We need to assert  that we shall get a set of linear equations
that is completely equivalent (order by order) to the original set without any approximation. Moreover, the linear set in our work is able to accommodate the non-perturbative data as well. In the following sections we apply the algorithm for different problems and for different type of series. The application of the algorithm will address first the weak-coupling parametrization and then will deal with cases of a mixture of information from weak-coupling, strong coupling and large-order data.
 
The structure of the prepare will be as follows. Sec.\ref{large-weak} addresses the high-order weak-coupling parametrization of a given series either with zero-radius of convergence or with a finite radius of convergence. In this section, with the aid of a relatively high order of the given series as input,  it has been shown how to get accurate predictions for the non-perturbative information like strong-coupling and large-order asymptotic behaviors.  In sec.\ref{strong-large}, we stress the high-order parametrization of the   hypergeometric approximants for divergent series with zero-radius of convergence using a mixture of information like weak-coupling, strong-coupling and large-order data. It will be shown in that section how non-perturbative data are able to accelerate the convergence of the algorithm. Sec.\ref{strong-large-f} is devoted to  the high order weak-coupling, strong coupling and large-order parametrization for a series with  finite radius of convergence while  summary and conclusions follow in sec.\ref{conc}. 
 \section{Weak-coupling high-order parametrization of the Hypergeometric
approximants}\label{large-weak}
Based on the large-order asymptotic behavior of a series, we select the suitable
hypergeometric approximant \cite{universal}. This large-order   asymptotic
behavior usually  takes the form:
\begin{equation}
 c_n\sim \alpha\left(  \left(  p-q-1\right)
n\right)  !(-\sigma)^{n}n^{b}\left(  1+O\left(  \frac{1}{n}\right)  \right),  
\end{equation}
which guides us to the suitable hypergeometric approximant out of the approximants:
	\[ _{p}F_{q}(a_{1},a_{2,}........a_{p};b_{1},b_{2},....b_{q};\sigma z). 
		\]
For instance, for a series with a finite radius of convergence or equivalently
$\left(  p-q-1\right)  =0$, the suitable approximant is $_{p}F_{p-1}%
(a_{1},a_{2,}........a_{p};b_{1},b_{2},....b_{p-1};\sigma z)$ while for a
series with zero radius of convergence and a large-order asymptotic behavior
of the form $\alpha n!(-\sigma)^{n}n^{b}\left(  1+O\left(  \frac{1}{n}\right)
\right)  $ or equivalently $\left(  p-q-1\right)  =1$, the suitable
approximant is $_{p}F_{p-2}(a_{1},a_{2,}........a_{p};b_{1},b_{2}%
,....b_{p-2};\sigma z)$. We will concentrate only on these types of divergent
series as the extension to the  other types is direct \cite{universal}.

\subsection{High-order parametrization of a divergent series with zero-radius
of convergence}

Consider a series for which we know the first $m+1$ terms as
\[
Q\left(  z\right)  \approx\sum_{0}^{m}c_{n}z^{n}.
\]
Assume that the series has a zero-radius of convergence with   an $n!$
growth factor in its large-order asymptotic behavior.  Accordingly, the
suitable hypergeometric approximant is $_{p}F_{p-2}(a_{1},a_{2,}%
........a_{p};b_{1},b_{2},....b_{p-2};\sigma z)$. This approximant is suitable in the sense that  it is the only type of hypergeometric functions that can be parametrized to give the same asymptotic behavior. For the  weak-coupling parametrization of a hypergeometric function, we have to solve the set of equations $R_{n}=\frac{c_{n}}{c_{n-1}}=g_{n}$ with $g_{n}$ is
given in Eq.(\ref{hnn}) or
\begin{equation}
R_{1}=g_{1},R_{2}=g_{2},.......R_{n}=g_{n}, \label{gna}%
\end{equation}
where $2p-1=m$. \ Note that  $m$ here is odd. For even $m$, one can use the once subtracted
series instead. From Eq.(\ref{hnn}), we have
\begin{equation}
g_{n}=\ \ \frac{%
{\displaystyle\prod_{i=1}^{p}}
\left(  a_{i}+n-1\right)  }{n%
{\displaystyle\prod_{j=1}^{p-2}}
\left(  b_{j}+n-1\right)  }\sigma. \label{gna-f} \nonumber
\end{equation}
Let us clarify the idea for $m=5$ or equivalently $p=3$. Then
\begin{equation}
g_{n}=\ \ \frac{%
{\displaystyle\prod_{i=1}^{3}}
\left(  a_{i}+n-1\right)  }{n\ \left(  b_{1}+n-1\right)  }\sigma.
\label{polydn}%
\end{equation}
$g_{n}$ can be rewritten in the form:%
\begin{align}
g_{n}  &  =\frac{%
{\displaystyle\prod_{i=1}^{3}}
\left(  a_{i}+n-1\right)  }{n\ \left(  b_{1}+n-1\right)  }\sigma\nonumber\\
&  =\frac{\sum_{i=0}^{i=3}d_{i}n^{i}}{\ \sum_{j=1}^{j=2}e_{j}n^{j}},\label{gnlin}
\end{align}
where $\ $
\begin{align}\label{lin-a-d}
d_{3}  &  =\sigma,\nonumber\\
d_{2}  &  =\sigma\left(  a_{1}+a_{2}+a_{3}-3\right) \nonumber\\
d_{1}  &  =\sigma\left(  \left(  a_{3}-1\right)  \left(  a_{1}+a_{2}-2\right)
+\left(  a_{1}-1\right)  \left(  a_{2}-1\right) \right)\nonumber\\
d_{0}  &  =\sigma\left(  a_{1}-1\right)  \left(  a_{2}-1\right)  \left(
a_{3}-1\right) \\
e_{2}  &  =1\nonumber\\
e_{1}  &  =\left(  b_{1}-1\right) \nonumber  
\end{align}
In the above set of equations, the relation between the coefficients $d_{i}$ of the polynomial
$\sum_{i=0}^{i=3}d_{i}n^{i}$ and $\left(  1-a_{i}\right)  $ \ are exactly the
same as the relation between coefficients of the polynomial and its roots
(known by Vieta's formulas). For instance, to get the values of the original
numerator ($a_{i}$) and denominator ($b_{i})$ parameters,  one can resort to
Vieta's formulas which is relating the roots of a polynomial to its
coefficients. For a polynomial of the form $P\left(  x\right)  =\sum_{k=0}%
^{m}f_{k}x^{k}$, according to Vieta's formulas, we have the roots $(r_{i})$ relations:%
\begin{align}
r_{1}+r_{2}+.......r_{m-1}+r_{m}  &  =-\frac{f_{m-1}}{f_{m}},\nonumber\\
r_{1}r_{2}+r_{1}r_{3}+.......+r_{1}r_{m}  &  =\frac{f_{m-2}}{f_{m}%
},..........\nonumber\\
r_{1}\ r_{2}\ .......r_{m-1}r_{m}  &  =\left(  -1\right)  ^{m}\frac{f_{0}%
}{f_{m}}.
\end{align}
The set   in Eq.(\ref{lin-a-d}) satisfies  Vieta's formulas for both
$\left(  1-a_{i}\right)  $ and $\left(  1-b_{i}\right)  $ as roots for the polynomials $\sum_{i=0}^{3}d_{i}n^{i}$ and $\sum
_{j=1}^{p-1}e_{j}n^{j}$, respectively. In fact, this is also true for any order of the polynomials $\sum_{i=0}^{p}d_{i}n^{i}$ and $\sum
_{j=1}^{p-1}e_{j}n^{j}$ with $e_{p-1}=1$ .
One can then solve the set of linear equations (linear in $d_{i}$
and $e_{j}$) of the form:
\begin{align}
d_{0}+d_{1}+d_{2}+d_{3}  &  =R_{1}\left(1+e_{1}\right) \nonumber\\
d_{0}+2d_{1}+4d_{2}+8d_{3}  &  =R_{2}\left(  4+2e_{1}\right) \nonumber\\
d_{0}+3d_{1}+9d_{2}+27d_{3}  &  =R_{3}\left( 9+ 3e_{1}\right) \nonumber\\
d_{0}+4d_{1}+16d_{2}+64d_{3}  &  =R_{4}\left( 16+ 4e_{1}\right) \label{lin-set}\\
d_{0}+5d_{1}+25d_{2}+125d_{3}  &  =R_{5}\left( 25+ 5e_{1}\right)  .\nonumber
\end{align}
This set has to be solved for the five unknowns $d_{0},d_{1},d_{2},d_{3},$
$e_{1}$ where the large order parameter $\sigma$   is equal to $d_{3}$.
The roots of the polynomials can be obtained as:
\begin{align}
\sum_{i=0}^{p}d_{i}\left(  1-A\right)  ^{i}  &  =0\nonumber\\
\sum_{j=1}^{p-1}e_{j}\left(  1-B\right)  ^{j-1}  &
=0\label{ndroots}
\end{align}
with the different roots are   $a_{i}$ and $b_{i}$ respectively. In the
following, we apply  this algorithm,  in which  the calculation are taking short time to obtain the high order parametrization for different type of series. Note that, solving the original set of non-linear equations (\ref{gna}) will take a relatively  long time to parametrize the $6th$ order using normal PC. The time needed increases non-linearly with the order and  might be impossible to solve a set of equations parametrizing the $25th$ order for instance.

\subsubsection{\texorpdfstring{$x^4$} \ anharmonic oscillator}

To test our formulas, let us consider the Hamiltonian of $x^{4}$ anharmonic
oscillator given by;%
\begin{equation}
H=\frac{p^{2}}{2}+\ \frac{1}{2}x^{2}+g x^{4}. \label{phi4}%
\end{equation}
The corresponding perturbation series for the ground state energy has
the form \cite{x4-bender-69}
\begin{equation}
E_{0}=\frac{1}{2}+\frac{3}{4}g-\frac{21}{8}g^{2}+\frac{333}{16}g^{3}%
-\frac{30885}{128}g^{4}+\frac{916731}{256}g^{5}\ +O(5^{6}). \label{x4-pert}%
\end{equation}
As we mentioned above, since this series is known to have a zero-radius of
convergence, it can be approximated by
\[
E_{0}=\frac{1}{2}\ {}_3F_{1}(a_{1},a_{2,}a_{3,};b_{1};\sigma z).
\]
From Eq.(\ref{non-linear}), we have the set of non-linear equations:%
\begin{align}
\ \frac{a_{1}a_{2}a_{3}}{b_{1}}\sigma &  =\frac{\frac{3}{4}}{\frac{1}{2}%
}\nonumber\\
\frac{a_{1}a_{2}a_{3}\left(  a_{1}+1\right)  \left(  a_{2}+1\right)  \left(
a_{3}+1\right)  }{2!b_{1}\left(  b_{1}+1\right)  }\sigma^{2}  &  =\frac
{-\frac{21}{8}}{\frac{1}{2}}\nonumber\\
\ \frac{a_{1}a_{2}a_{3}\left(  a_{1}+1\right)  \left(  a_{2}+1\right)  \left(
a_{3}+1\right)  \left(  a_{1}+2\right)  \left(  a_{2}+2\right)  \left(
a_{3}+2\right)  }{3!b_{1}\left(  b_{1}+1\right)  \left(  b_{1}+2\right)
}\sigma^{3}  &  =\frac{\frac{333}{16}}{\frac{1}{2}}\nonumber\\
\ \frac{a_{1}a_{2}a_{3}\left(  a_{1}+1\right)  \left(  a_{2}+1\right)  \left(
a_{3}+1\right)  ......\left(  a_{1}+3\right)  \left(  a_{2}+3\right)  \left(
a_{3}+3\right)  }{4!b_{1}\left(  b_{1}+1\right)  ....\left(  b_{1}+3\right)
}\sigma^{4}  &  =\frac{-\frac{30885}{128}}{\frac{1}{2}}\label{non-lin-num}\\
\ \frac{a_{1}a_{2}a_{3}\left(  a_{1}+1\right)  \left(  a_{2}+1\right)  \left(
a_{3}+1\right)  ......\left(  a_{1}+4\right)  \left(  a_{2}+4\right)  \left(
a_{3}+4\right)  }{5!b_{1}\left(  b_{1}+1\right)  ....\left(  b_{1}+4\right)
}\sigma^{5}  &  =\frac{\frac{916731}{256}}{\frac{1}{2}}\nonumber
\end{align}
The solution of this set gives the values $a_{1}=0.364464,a_{2}%
=-0.335298,a_{3}=\ 5.89769,b_{1}=1.16276,\sigma=-2.41998$. Note that any
permutation between the numerator or the denominator parameters will leave the
hypergeometric approximant the same. Thus the other solutions to the
non-linear set are all equivalent. Now, let us solve for these parameters but in
using the equivalent linear set in Eq.(\ref{gnlin}) as%
\begin{equation}
\frac{\sum_{i=0}^{i=3}d_{i}n^{i}}{n^{2}+ e_{1}n}=R_{n},
\end{equation}
or
\begin{equation}
d_{0}+d_{1}n+d_{2}n^{2}+d_{3}n^{3}=R_{n}\left(  n^{2}+e_{1}n\right)
\end{equation}%
Explicitly we have:
\begin{align}
d_{0}+d_{1}+d_{2}+d_{3}  &  =\frac{\frac{3}{4}}{\frac{1}{2}}\left(
e_{1}+1\right) \nonumber\\
d_{0}+2d_{1}+4d_{2}+8d_{3}  &  =2\frac{-\frac{21}{8}}{\frac{3}{4}}\left(
e_{1}+2\right) \nonumber\\
d_{0}+3d_{1}+9d_{2}+27d_{3}  &  =3\frac{\frac{333}{16}}{-\frac{21}{8}}\left(
e_{1}+3\right) \nonumber\\
d_{0}+4d_{1}+16d_{2}+64d_{3}  &  =4\frac{-\frac{30885}{128}}{\frac{333}{16}%
}\left(  e_{1}+4\right) \\
d_{0}+5d_{1}+25d_{2}+125d_{3}  &  =5\frac{\frac{916731}{256}}{-\frac
{30885}{128}}\left(  e_{1}+5\right) \nonumber
\end{align}
we get the values $e_{1}=0.162756,d_{0}=-10.0582,d_{1}=21.3053,d_{2}%
=-7.08295,d_{3}=-2.41998$. Here $d_{3}=\sigma=-2.41998$ which is the same
result we obtained above by using the direct non-linear set of equations. To
get the numerator parameters, we find the roots of the polynomial:%

\begin{equation}
-10.0582+21.3053\left(  1-A\right)  -7.08295\left(  1-A\right)  ^{2}%
-2.41998\left(  1-A\right)  ^{3}=0,
\end{equation}
which has the roots $a_{1}=-0.335298$, $a_{2}=0.364464$ $,a_{3}=5.89769$. These are   the same results we obtained from the solution of the non-linear set Eq.(\ref{non-lin-num}). For the denominator parameter $b_{1},$ it can be found
from the roots of the polynomial (in this case only one root)
\[
(1-B)+0.162756\ =0,
\]
which gives the result $b_{1}=1.16276$ . Again, it is the same result obtained
from the solution of the set in Eq.(\ref{non-lin-num}). So the recipe followed
by solving first the linear set
\begin{equation}
\sum_{i=0}^{i=p}d_{i}n^{i}=R_{n}\left(  n^{p-1}+\sum_{j=1}^{p-2}e_{j}%
n^{j}\right)  ,\ \ e_{p-1}=1 \label{lin-d-e}%
\end{equation}
then getting all the parameters from solving the polynomial equations
($d_{p}=\sigma$)%

\begin{align}
\sum_{i=0}^{i=p}d_{i}\left(  1-A\right)  ^{i}  &  =0\text{, with }a_{i}\text{
are the roots,}\nonumber\\
 \sum_{j=1}^{p-1}e_{j}\left(  1-B\right)  ^{j-1}  &
=0\text{, with }b_{i}\text{ are the roots,} \label{linear-a-b}%
\end{align}
 is equivalent (order by order) to solve the actual set of non-linear equations. However, using
Eqs.(\ref{lin-d-e},\ref{linear-a-b} ) is much faster and up to high orders of 
calculations can be carried out in short time. Note that the exact value for
the large order parameter $\sigma$ is $-3$ \cite{harm-sigma} while our fifth
order prediction is $d_{3}=\sigma=-2.41998$. As we will see later, the
prediction of $\sigma$ will be improved greatly as we increase the order of
the used perturbation series. Moreover, the asymptotic strong coupling
behavior is known to take the value $s^{\ast}=-\min\{\operatorname{Re}%
(a_{i})\}$ \cite{Analytic2016} or in other words for large $g$ we have: 
\begin{equation}
E_{0}\propto g^{s^{\ast}} \label{strong-ss}.
\end{equation}
Our fifth order prediction is $s^{\ast
}=0.335298$ while the exact result is $s^{\ast}=\frac{1}{3}\approx0.333333$
\cite{har4}. Needles to say that we have used the relation in
Eq.(\ref{hyp-G-C}) for the analytic continuation of the hypergeometric
approximants to the region $g>0.$ For instance, the fifth order parametrization
of the ground state energy $E_{0}$ is
\begin{equation}
E_{0}\left(  g\right)  \simeq_{\text{ }3}F_{1}(a_{1},a_{2},a_{3}%
;b_{1}\ ;\sigma z)=\frac{ \Gamma\left(  b_{1}\right)  }%
{\prod_{k=1}^{3}\Gamma\left(  a_{k}\right)  }\MeijerG*{1}{p}{p}{q+1}{1-a_{1}%
,1-a_{2},1-a_{3}}{0,1-b_{1} }{\sigma g},
\end{equation}
which leads to the results $E_{0}\left(  0.5\right)  \simeq0.696241$ compared
to the exact result $0.6961768$ from Ref.\cite{har4} and
$  E_{0}\left(  1\right)  \simeq0.804008$ compared to the
exact result $0.803771$ from the same reference. Let us \ consider the 25th order
parametrization for $E_{0}\left(  g\right)  $ where our parametrization gives
the result $E_{0}\left(  2\right)  \simeq0.9515684743$ while the exact result
$0.9515684727$ from Ref.\cite{har4}. Our results shares the first 8 digits
with the exact result. 
In table \ref{x4w}, we list the hypergeometric approximation for different orders  of the series representing  $E_{0}\left(  g\right)$ which shows  convergence to exact values as order increases. Note that at some orders
(especially high ones) we find singularities in the approximants which means
that such orders need to be skipped to other ones   that have no such singularities.
\begin{table}[H]
\centering 
\caption{{\protect\scriptsize { Comparison between  our prediction for the ground state energy of the  $x^4$ anharmonic oscillator and numerical results $E_{exact}$ from Ref.\cite{har4}. One can see from the table that the results are greatly improved as we increase the size of weak-coupling information (increasing the order). Here, we included only odd orders however even orders can be approximated by considering the once-subtracted series.}}}%
\label{x4w}%
\begin{tabular}{|l|l|l|l|l|l|l|l|l|}
\hline
g   & \begin{tabular}[c]{@{}l@{}}\ \ ${}_2F_0$\\\  3rd order\end{tabular} & \begin{tabular}[c]{@{}l@{}}\ \ ${}_3F_1$\\\ 5th order\end{tabular} & \begin{tabular}[c]{@{}l@{}}\ \ ${}_5F_3$\\\ 9th order\end{tabular} & \begin{tabular}[c]{@{}l@{}}\ \ ${}_8F_6$\\\ 15th order\end{tabular} & \begin{tabular}[c]{@{}l@{}}\ \ ${}_{12}F_{10}$\\\ 23rd order\end{tabular} & \begin{tabular}[c]{@{}l@{}}\ \ ${}_{13}F_{11}$\\\ 25th order\end{tabular} & \begin{tabular}[c]{@{}l@{}}\ \ ${}_{15}F_{13}$\\\ 29th order\end{tabular} & Exact    \\ \hline
0.1 & 0.559029                                                 & 0.559147                                                & 0.559146                                                & 0.559146                                                 & 0.559146                                                   & 0.559146                                                   & 0.559146                                                   & 0.559146 \\ \hline
0.5 & 0.692890                                                 & 0.696241                                                & 0.696174                                                & 0.696176                                                 & 0.696176                                                   & 0.696176                                                   & 0.696176                                                   & 0.696176 \\ \hline
1   & 0.794363                                                 & 0.804008                                                & 0.803755                                                & 0.803771                                                 & 0.803771                                                   & 0.803771                                                   & 0.803771                                                   & 0.803771 \\ \hline
2   & 0.928912                                                 & 0.95224                                                 & 0.951499                                                & 0.951568                                                 & 0.951568                                                   & 0.951568                                                   & 0.951568                                                   & 0.951568 \\ \hline
50  & 2.15034                                                  & 2.51369                                                 & 2.49463                                                 & 2.49964                                                  & 2.49969                                                    & 2.49971                                                    & 2.49971                                                    & 2.49971  \\ \hline
\end{tabular}
\end{table}

What is really more impressive is to get accurate predictions for the parameters characterizing the asymptotic behavior of the given series from just weak-coupling (perturbation) information. Our 25th order parametrization gives $\sigma\simeq-2.9999550507564794$ compared to
the exact value $\sigma=-3$ while our 25th prediction for $s^{\ast}%
\simeq0.333337965$ which shares the first $5$ digits with the exact value
$s^{\ast}=1/3\simeq0.333333$. Moreover, the asymptotic large order behavior
for the series $E_{0}\left(  g\right)  =\sum_{n=0}^{\infty}c_{n}g^{n\text{ }}$
is known to take the form $c_{n\rightarrow\infty}\propto n!n^{b}\sigma^{n}$.
Our prediction for $b$ can be calculated from the relation
\[
\sum_{i=1}^{p}a_{i}-\sum_{i=1}^{q}b_{i}-2=b,
\]
where our 25th order prediction gives $b\simeq-0.492309$ compared to the well
known exact result $b\simeq-\frac{1}{2}$ \cite{har4}.

In table \ref{x4wp}, we list our hypergeometric prediction for the non-perturbative parameters $S^*$, $\sigma$ and $b$ and compare them to their exact values \cite{harm-sigma}. Specially for the strong coupling parameter  $S^*$ and the large-order parameter $\sigma$ one can get good approximation with relatively small number of terms from the weak-coupling expansion as an input. The predictions  are improved by adding more terms as input. The results extracted from approximants parametrized from relatively high orders are shown to be very accurate. However, accurate and stable results for  the large-order parameter $b$ from weak-coupling information can only be extracted using relatively high-orders as can be seen from the table. A note to be mentioned is that although for some orders like the 31st order in the table, the hypergeometric approximant is singular but it gives accurate results for the non-perturbative parameters. The point is that while for singular cases the hypergeometric approximant represents accurately the given series, the analytic continuation conditions from hypergeometric approximant to the Meijer-G functions \cite{Analytic2016}  are not satisfied.    

\begin{table}[pth]
\caption{{\protect\scriptsize { In this table we list the hypergeometric prediction for the strong coupling parameter $S^*$ and the large-order parameters $\sigma$ and $b$. We compare our predictions to the well known exact results \cite{harm-sigma}. High orders give accurate results for the mentioned parameters..}}}%
\label{x4wp}%
\begin{tabular}{|l|l|l|l|l|l|l|l|l|}
\hline
Parameter & \begin{tabular}[c]{@{}l@{}}\ \ ${}_2F_0$\\  3rd order\end{tabular} & \begin{tabular}[c]{@{}l@{}}\ \ ${}_3F_1$\\\ 5th order\end{tabular} & \begin{tabular}[c]{@{}l@{}}\ \ ${}_5F_3$\\\ 9th order\end{tabular} & \begin{tabular}[c]{@{}l@{}}\ \ ${}_8F_6$\\\ 15th order\end{tabular} & \begin{tabular}[c]{@{}l@{}}\ \ ${}_{12}F_{10}$\\\ 23rd order\end{tabular} & \begin{tabular}[c]{@{}l@{}}\ \ ${}_{13}F_{11}$\\\ 25th order\end{tabular} & \begin{tabular}[c]{@{}l@{}}\ \ ${}_{16}F_{14}$\\\ 31st order\end{tabular} & Exact    \\ \hline
$\ \ \ S^*$     & 0.273262                                                 & 0.335298                                                & 0.331019                                                & 0.333228                                                 & 0.333299                                                   & 0.333338                                                   & 0.333327                                                   & 0.333333 \\ \hline
$\ \ \ \sigma$  & -4.14286                                                 & -2.41998                                                & -3.02560                                                & -2.94774                                                 & -3.00106                                                   & -2.99996                                                   & -3.00010                                                   & -3.0     \\ \hline
\ \ \ b         & -0.273262                                                & 3.33691                                                 & -0.114857                                               & 1.35876                                                  & -0.584225                                                  & -0.492309                                                  & -0.513786                                                  & -0.5     \\ \hline
\end{tabular}
\end{table}

\subsubsection{\texorpdfstring{$ix^{3}\ \mathcal{PT}-$}\ symmetric anharmonic oscillator}

For the Hamiltonian  of the form:%
\begin{equation}
H=\frac{1}{2}p^{2}+\frac{1}{2}m^{2}x^{2}\ +\frac{i\sqrt{g}}{6}x^{3},
\end{equation}
The ground state energy up to the 20th order has been obtained in
Ref.\cite{ix3-bender-large} as :
\begin{equation}
E_{0}=\frac{1}{2}+\frac{11g}{288}-\frac{930}{288^{2}}g^{2}+\frac
{158836}{288^{3}}g-\frac{38501610}{288^{4}}g^{4}+O\left(  g^{5}\right)  .
\label{pertub}%
\end{equation}
This series   has a zero radius of convergence and its coefficients have an
asymptotic large-order behavior of the form $\alpha n!\sigma^{n}n^{b}\left(
1+O\left(  \frac{1}{n}\right)  \right)  $ with $\sigma=-\frac{5}{24}$ and
$b=-\frac{1}{2}$ \cite{ix3-bender-large}. According to the above discussions,
the suitable hypergeometric approximants are then $_{p}F_{p-2}(a_{1},a_{2,}........a_{p};b_{1},b_{2},....b_{p-2};\sigma g)$.
The lowest order weak-coupling approximant $_{2}F_{0}\left(  a_{1},a_{2};\text{ };\sigma g\right)$ has
three unknowns $a_{1},a_{2}\ $and $\sigma$ to be determined using the set of
equations:%
\begin{align}
d_{0}+d_{1}+d_{2}\  &  =R_{1}\ ,\nonumber\\
d_{0}+2d_{1}+4d_{2}\  &  =2R_{2}\ ,\nonumber\\
d_{0}+3d_{1}+9d_{2}\  &  =3R_{3}\ .
\end{align}
Here
\begin{equation}
R_{1}=\frac{\frac{11}{288}}{\frac{1}{2}},\text{ }R_{2}=\frac{-\frac
{930}{288^{2}}}{\frac{11}{288}},R_{3}=\frac{\frac{158836}{288^{3}}}%
{-\frac{930}{288^{2}}}%
\end{equation}
which gives
\[
d_{0}=\frac{12979}{61380},d_{1}=\frac{31711}{245520}\ ,d_{2}=-\frac{901}%
{3410}.\
\]
Then one can get the values of the numerator parameters $a_{1}$ and $a_2$ from the
following root equation:
\[
\sum_{i=0}^{i=p}d_{i}\left(  1-A\right)  ^{i}=0,
\]
which gives%
\begin{equation}
a_{1}=\frac{98033-\sqrt{14477166529}}{129744},\text{ }a_{2}=\frac
{98033+\sqrt{14477166529}}{129744}.
\end{equation}
Note that as shown in the $x^{4}$ case discussed above, our third order
approximation for the parameter $\sigma$ is given by $d_{2}=-\frac{901}%
{3410}\approx$ $-0.264\,22$. In fact, the exact value of $\sigma$ is given by
$\frac{-5}{24}=\allowbreak-0.208\,33$. The corresponding approximant is given
by:
\[
E_{0}\left(  g\right)  \approx\frac{1}{2}\ {}_2F_{0}(a_{1},a; \ ;\sigma g).
\]
This third order approximant results in $E_{0}\left(  1\right)  =0.530752$
compared to exact value of $0.530782$ from Ref.\cite{Zinnx3}. In table \ref{x3w}, we
list the different orders approximants up to O(19) and compare to the numerical
results from Ref. \cite{ix3-bender-large}. Needless to say that the algorithm gives accurate predictions as the order increases. In this table there is an empty cell for the 19th order at the very small coupling value $0.0070313$ . In fact, this happens for other approximants as well (not shown) and the reason is that the point  $g=0$ is a regular singular point of the Meijer-$G$ function. In general, this affects the accuracy of the results for the approximants near $g=0$.

For the prediction of the non-perturbative parameters from the hypergeometric approximants for the weak-coupling series, the strong coupling parameter $S^*$ is predicted to be $0.171785$ for ${}_2F_0$ and get improved to $0.199960$ for ${}_{10}F_8$ while the exact value is given by $S^*=0.2$ \cite{Zinnx3}. Also, for the parameter $\sigma$ we get the result $\sigma=-0.264223$ from the approximant ${}_2F_0$ and more accurate prediction is obtained from the higher order approximant ${}_{10}F_8$ where we get $\sigma=-0.207996$. The exact value is $\sigma=\frac{5}{24}\approx0.20833$\cite{Zinnx3}. The large-order parameter $b$, on the other hand, needs a relatively high order of perturbative terms as input and the value predicted by the $19th$ order ( not high enough)  approximant is $-0.183128$ compared to the exact value of $-\frac{1}{2}$ \cite{Zinnx3}.
\begin{table}[H]
\centering
\caption{{\protect\scriptsize {  Our hypergeometric approximants for the ground state energy of the $\mathcal{PT}$-symmetric $ix^3$ theory up to 19th order of the input perturbation series. The results are compared to numerical calculations from Ref.\cite{ix3-bender-large} (with our coupling $g$ is related to their coupling $\lambda$ as $g\equiv 288\lambda^2$). The empty cell means that the approximant is singular for this coupling value.}}}%
\label{x3w}%
\begin{tabular}{|l|l|l|l|l|l|}
\hline
g         & \begin{tabular}[c]{@{}l@{}}\ \ ${}_2F_0$\\  3rd order\end{tabular} & \begin{tabular}[c]{@{}l@{}}\ \ ${}_4F_2$\\ 7th order\end{tabular} & \begin{tabular}[c]{@{}l@{}}\ \ ${}_7F_5$\\ 13th order\end{tabular} & \begin{tabular}[c]{@{}l@{}}\ \ ${}_{10}F_8$\\ 19th order\end{tabular} & Exact   \\ \hline
0.0070313 & 0.50263                                                  & 0.50263                                                 & 0.50263                                                  &                                                           & 0.50263 \\ \hline
0.28125   & 0.50998                                                  & 0.50998                                                 & 0.50998                                                  & 0.50995                                                   & 0.50998 \\ \hline
1. 125    & 0.53389                                                  & 0.53393                                                 & 0.53393                                                  & 0.53393                                                   & 0.53393 \\ \hline
4.5       & 0.59408                                                  & 0.59491                                                 & 0.59492                                                  & 0.59492                                                   & 0.59492 \\ \hline
18        & 0.70660                                                  & 0.71290                                                 & 0.712935                                                 & 0.71294                                                   & 0.71294 \\ \hline
72        & 0.87574                                                  & 0.90002                                                 & 0.90025                                                  & 0.90026                                                   & 0.90026 \\ \hline
288       & 1.10343                                                  & 1.16652                                                 & 1.16737                                                  & 1.16745                                                   & 1.16746 \\ \hline
1152      & 1.39749                                                  & 1.52823                                                 & 1.53047                                                  & 1.53074                                                   & 1.53078 \\ \hline
\end{tabular}
\end{table}

\subsection{High-order parametrization of a series with finite radius of convergence}

Our Hypergeometric-Meijer approximation algorithm is a generalized one in the sense that it is not only approximating series with zero radius of convergence, but also can analytically continue a sires with finite radius of convergence to values outside the disk of convergence.   In certain situations in physics, one can find different important cases where the perturbation series has
a finite radius of convergence but the region of interest lies outside the disk of convergence. In this case, one needs to find an algorithm capable to extend the approximation to values beyond the radius of convergence.
To do that, one can take into account the fact that for such type of series,
the asymptotic large-order behavior takes the form:
\[
c_{n}\sim\alpha\ (-\sigma)^{n}n^{b}\left(  1+O\left(  \frac{1}{n}\right)
\right)  ,
\]
where $c_{n}$ is the $n^{th}$ coefficient of the perturbation series. According
to our approximation recipe, the approximant $_{p}F_{p-1}(a_{1},a_{2,}%
........a_{p};b_{1},b_{2},....b_{p-1};\sigma z)$ is the suitable one for such
series. This is  because the hypergeometric approximant and the given series possess the same
form of the asymptotic large-order behavior. In other words, this hypergeometric approximant can be parametrized to account for all the features that the given series has.  In the following, we shall
consider two different series of such type and show that the  hypergeometric
approximants can give  very accurate results. Note that, the hypergeometric approximant $_{p}F_{p-1}(a_{1},a_{2,}%
........a_{p};b_{1},b_{2},....b_{p-1};\sigma z)$ has a branch cut in the $\sigma z$ interval  $\{1,\infty\}$. Certain tricks are thus needed to approximate a series with non-alternating signs of the series coefficients. A problem that resembles the non-Borel summability in the Borel resummation method. We shall see that one can overcome it with some tricks. 

\subsubsection{Strong-coupling expansion of the \texorpdfstring{$ix^{3}$}\  theory (Yang-Leemodel)}

As an example for a series with finite-radius of convergence, we consider the ground state-energy of the one dimensional Yang-Lee model where the Hamiltonian takes the from:%
\begin{equation}
H_{J\text{ }}=\frac{p^{2}}{2}+\frac{ix^{3}}{6}+\frac{1}{2}iJx. \label{Hamilt}%
\end{equation}
In fact, the weak-coupling ($J$) expansion  of  this Hamiltonian represents  the strong-coupling ($g$) expansion of    the $ix^{3}$ $\mathcal{PT}-$symmetric anharmonic oscillator \cite{Zinnx3}. Note that, this
model is important toward the study of the Lee-Yang-Edge singularity
\cite{Young-Lee1,Yang-Lee2}. The perturbation series up to $O(20)$ in $J$
 for the ground state energy has been obtained in Ref. \cite{Zinnx3} (Eq.(92)
there, with $J\equiv\chi$):%
\begin{align}
E_{0}^{J}  &  =\sum_{n=0}^{\infty}c_{n}J^{n}%
=0.3725457904522070982506011+0.3675358055441936035304J\label{Jpert}\\
&  +0.1437877004150665158339J^{2}+0.0265861056270593871352J^{3}+(O\left(
J^{4}\right)  .\label{EJ}
\end{align}
In this case $g_{n}$ in Eq.(\ref{hnn}) reduces to:%
\begin{equation}
g_{n}=\ \ \frac{%
{\displaystyle\prod_{i=1}^{p}}
\left(  a_{i}+n-1\right)  }{n%
{\displaystyle\prod_{j=1}^{p-1}}
\left(  b_{j}+n-1\right)  }\sigma.
\end{equation}
To obtain the parameters in the approximant $_{p}F_{p-1}(a_{1},a_{2,}%
........a_{p};b_{1},b_{2},....b_{p-1};\sigma g)$, one has to solve the set of
$2p$ equations of the form:
\begin{equation}
\frac{%
{\displaystyle\prod_{i=1}^{p}}
\left(  a_{i}+n-1\right)  }{n%
{\displaystyle\prod_{j=1}^{p-1}}
\left(  b_{j}+n-1\right)  }\sigma=\frac{c_{n}}{c_{n-1}}, \label{hypset}%
\end{equation}
where $c_n$ is the $n^{th}$ coefficient in the above series. This set  is a non-linear one and for high order parametrization it will take a very long time to solve which make the process impractical for such high orders. The approximant $_{p}F_{p-1}%
(a_{1},a_{2,}........a_{p};b_{1},b_{2},....b_{p-1};\sigma g)$ has been used
as Borel functions in Ref.\cite{hyper-borel} where the authors introduced the ansatz in
Eq.(5) in the same reference to obtain a linear set of equations. In our work,
we will not employ this ansatz and instead we will try to   get a  set of linear equations     that is  exactly and order by order equivalent to the set in Eq.(\ref{hypset}). The left hand side of Eq.(\ref{hypset}) can be written in
the form:
\begin{equation}
g_{n}=\frac{\sum_{i=0}^{p}d_{i}n^{i}}{\ \sum_{j=1}^{p}e_{j}n^{j}},
\end{equation}
with $d_{p}=\sigma$ and $e_{p}=1$. Let us elucidate it for the parametrization of
the fourth order approximant $_{2}F_{1}(a_{1},a_{2};b_{1};\sigma g)$. In this
case $p=2$ and the set in Eq.(\ref{hypset}) reduces to the linear set:%

\begin{equation}
\sum_{i=0}^{p}d_{i}n^{i}=\frac{c_{n}}{c_{n-1}}\sum_{j=1}^{p}e_{j}n^{j}
\label{linsetp1}%
\end{equation}
\begin{align}
\ d_{0}+d_{1}+d_{2}\  &  =\frac{0.3675358055441936035304}%
{0.3725457904522070982506011}\left(  1+e_{1}\right) \nonumber\\
d_{0}+2d_{1}+4d_{2}  &  =\ \frac{0.1437877004150665158339}%
{0.3675358055441936035304}\left(  4+2e_{1}\right) \nonumber\\
d_{0}+3d_{1}+9d_{2}  &  =\ \frac{-0.0265861056270593871352}%
{0.1437877004150665158339}\left(  9+3e_{1}\right) \\
d_{0}+4d_{1}+16d_{2}  &  =\ \frac{0.0098871650792008872905}%
{-0.0265861056270593871352}\left(  16+4e_{1}\right) \nonumber
\end{align}
which gives $e_{1}=-1.\,\allowbreak788\,9,d_{0}=-3.\,\allowbreak
502\,2,d_{1}=3.\,\allowbreak614\,1,d_{2}=-0.890\,21$. Note that the large
order parameter $\sigma$ is in our calculation equal to $d_{2}$. Accordingly,
the fourth order hypergeometric approximant for the critical coupling $J_{c}$
is $\frac{1}{\sigma}=\frac{1}{-0.890\,21}=\allowbreak-1.\,\allowbreak
123\,3\ $\ compared to $-1.351$ from Ref.\cite{Zinnx3} . Of course increasing
the order will greatly improve this   prediction as we will see. Back to
the parameters $a_{i}$ and $b_{i}$ in the hypergeometric approximant $_{2}%
F_{1}(a_{1},a_{2};b_{1};\sigma g)$. For $a_{1},a_{2}$, they represents the
roots of the polynomial equation:%
\begin{align}
\sum_{i=0}^{2}d_{i}\left(  1-A\right)  ^{i}  &  =0,\nonumber\\
-0.890\,21\left(  1-A\right)  ^{2}+3.\,\allowbreak614\,1\left(  1-A\right)
-3.\,\allowbreak502\,2  &  =0,
\end{align}
which gives
\[
a_{1}=-0.59814,\ a_{2}=-1.4617\allowbreak.
\]
Likewise, to get $e_1$ we solve the equation 
\begin{equation}
 \sum_{j=1}^{p}e_{j}\left(  1-B\right)  ^{j-1}  =0.
\end{equation}
In this case we solve $(1-B)+e_1=0$, which gives $b_1=e_1+1=-0.78891$.
 As we said, in our algorithm the set of linear equations (Eq.(\ref{linsetp1})) is
equivalent (order by order) to the actual non-linear set of equations in
Eq.(\ref{hypset}). One can double check by solving directly the non-linear set of equations:
\begin{align*}
\frac{a_{1}a_{2}\sigma}{b_{1}}  &  =\frac{0.3675358055441936035304}%
{0.3725457904522070982506011},\\
\frac{\left(  a_{1}+1\right)  \left(  a_{2}+1\right)  \sigma}{2\left(
b_{1}+1\right)  }  &  =\frac{0.1437877004150665158339}%
{0.3675358055441936035304}\ ,\\
\frac{\left(  a_{1}+2\right)  \left(  a_{2}+2\right)  \sigma}{3\left(
b_{1}+2\right)  }  &  =\frac{-0.0265861056270593871352}%
{0.1437877004150665158339}\ ,\ \\
\frac{\left(  a_{1}+3\right)  \left(  a_{2}+3\right)  \sigma}{4\left(
b_{1}+3\right)  }  &  =\frac{0.0098871650792008872905}%
{-0.0265861056270593871352}\ \ ,
\end{align*}
which gives the same results we obtained using our linear set above. Note that at a
relatively high order (seventh for instance) it would be very hard to solve
the non-linear set using a normal computer but it takes a normal PC just
seconds to solve the equivalent linear set of equations. Note also that, using
the ansatz in Eq.(5)$\allowbreak$ in Ref.\cite{hyper-borel} will not lead to
the same results as the ansatz there is an approximation that works better for
high orders. 

The fourth order  approximation gives: 
\[
E_{0}^{J}=0.372545790452207098250601\ _{2}F_{1}(-0.598,-1.\,\allowbreak
4617;-0.7889;-0.89021g).
\]
Our fourth order result for $J=-1$ \ gives $E_{0}^{J}=0.197526\ $  compared to
$0.195751$ of ODM method at a transformation order 150 \cite{Zinnx3} .

Table \ref{x3-lee-w-E} show our hypergeometric  approximation results up to $O(20)$.
Also, our prediction for the non-perturbative parameters ( from weak-coupling only as input) is shown in table \ref{x3-lee-w-p}. One can easily see the very accurate results compared to the well known  exact ones ( for $b$ and $S^*$) and the $150th$ order of the methods in Ref. \cite{Zinnx3} for $J_C$. Note that, the exact value of $S^*$ is known to be $\frac{3}{2}$ from Ref. \cite{Zinnx3} while the exact $b$ value is known to be $-\frac{3}{2}$ from Ref.\cite{Large-strong}. 
\begin{table}[H]
\centering
\caption{{\protect\scriptsize {Our predictions for the ground state energy of the model in Eq.(\ref{Hamilt})   compared to results from Continued Fraction method ($150$th order) in Ref.\cite{Zinnx3}.}}}
\label{x3-lee-w-E}%
\begin{tabular}{|c|llll|}
\hline
\multicolumn{1}{|l|}{}  & \multicolumn{4}{c|}{$E_0^J$}                                                                                                                                                                                                                                                             \\ \hline
\multicolumn{1}{|l|}{\ \ \ \ J} & \multicolumn{1}{c|}{\begin{tabular}[c]{@{}c@{}}${}_2F_1$ \\  4th order\end{tabular}} & \multicolumn{1}{c|}{\begin{tabular}[c]{@{}c@{}}${}_5F_4$ \\ 10th order\end{tabular}} & \multicolumn{1}{c|}{\begin{tabular}[c]{@{}c@{}}${}_{10}F_9$ \\ 20th order\end{tabular}} & \multicolumn{1}{c|}{Continued Fraction} \\ \hline
$-2^{4/5}$              & \multicolumn{1}{l|}{0.395189 -0.298946i}                                      & \multicolumn{1}{l|}{0.389793-0.363668i}                                       & \multicolumn{1}{l|}{0.3898 - 0.3644 i}                                       & 0.3898(5)-0.3644(3) i                   \\ \hline
$-5^{-4/5}$             & \multicolumn{1}{l|}{0.282699573003304}                                        & \multicolumn{1}{l|}{0.282699258193271}                                        & \multicolumn{1}{l|}{0.282699258188 }                                        & 0.2826992581932749098990(1)             \\ \hline
-1                      & \multicolumn{1}{l|}{0.197526449159134}                                        & \multicolumn{1}{l|}{0.1957508231732275}                                       & \multicolumn{1}{l|}{0.195750815711}                                       & 0.195750 8157161719(6)                  \\ \hline
$-21.6^{-4/5}$          & \multicolumn{1}{l|}{0.3421580192691438}                                       & \multicolumn{1}{l|}{0.3421580186193393}                                       & \multicolumn{1}{l|}{0.342158018619340}                                        & 0.34215801861934042140767(6)            \\ \hline
\end{tabular}
\end{table}
\begin{table}[H]
\centering
\caption{{\protect\scriptsize {The non-perturbative parameters predicted from our hypergeometric approximation for the series in Eq.(\ref{EJ}) are listed. The parameters are, the large-order parameter $b$ which is related to the critical exponent \cite{universal}, the critical coupling $J_C=-1/ \sigma$ and    the large-$J$ asymptotic parameter $S^*$ .}}}
\label{x3-lee-w-p}%
\begin{tabular}{|l|l|l|l|}
\hline
Approximant & \ \ \ \  b             & \ \ \ \ $J_c$         &\ \ \  \ $S^*$             \\ \hline
$\ \ \ \ {}_2F_1$         & -2.270915594  & -1.123334723 & 1.461687589   \\ \hline
$\ \ \ \ {}_3F_2$        & -1.196395939  & -1.369308564 & 1.499820990   \\ \hline
$\ \ \ \ {}_4F_3$         & -1.1910447654 & -1.374864586 & 1.499724548  \\ \hline
$\ \ \ \ {}_5F_4$         & -1.584244428  & -1.348955815 & 1.499960911  \\ \hline
$\ \ \ \ {}_6F_5$       & -1.448166413  & -1.351805067 & 1.500002881   \\ \hline
$\ \ \ \ {}_{10}F_9$      & -1.499872281  & -1.351039990 & 1.49999963   \\ \hline
\end{tabular}
\end{table}

\subsubsection{The high-temperature expansion for the susceptibility of the SQ Ising model}
Another example for a series of finite radius of convergence is the high temperature series expansion of the Ising model. The high temperature expansion (strong-coupling) is one of the powerful techniques to study critical phenomena   \cite{HT1,HT2}. In literature, one can find that the corresponding perturbation series is known up to a relatively
high order and thus expediting the calculation within the hypergeometric
approximation is more than important. To select the the suitable hypergeometric
approximants for such type of series, one has to take into account  that such series has a finite
radius of convergence and thus the approximants $_{p}F_{p-1}(a_{1}%
,a_{2,}........a_{p};b_{1},b_{2},....b_{p-1};\sigma \beta)$ are suitable and
expected to give accurate results.

The series of the susceptibility of the SQ (spin-half) Ising model is given in
Ref.\cite{HT1} up to $O(\beta^{25})$:%
\begin{align}
\chi\left(  \beta\right)   &  =1+4\beta+12\beta^{2}+\frac{104}{3}\beta{{}^{3}%
}+\dots\dots\dots+\frac{4735391065845611373232}{14992036723125}\beta{{}^{2}%
}{{}^{1}}\nonumber\\
&  +\frac{529562920319138348552816}{714620417135625}\beta{{}^{2}}{{}^{2}%
}+\frac{85616154520095267692857616}{49308808782358125}\beta{{}^{2}}{{}^{3}%
}\nonumber\\
&  +\frac{66773068948180944546678128}{16436269594119375\ }\beta^{24}%
+\frac{3192145249472459217984684656}{336196423516078125\ }\beta^{25}%
+\dots\dots\dots,\label{HT}
\end{align}
Here $\chi\left(  \beta\right)  $ is the susceptibility while $\beta$
represents the inverse temperature . Near the tip of the branch cut, ${\sigma
z=1,}$the approximants $_{\text{ }p}F_{p-1}\left(  {a_{1},......a_{p}%
;b_{1},........b_{q};\sigma z}\right)  $ are known to a have a power-law
behavior in the form \cite{HTF,Math,Bos-Hub,Bos-Hub1}: 
\begin{equation}
_{\text{ }p}F_{p-1}\left(  {a_{1},..a_{p};b_{1},..b_{p-1};\sigma z}\right)
-{}_{p}F_{p-1}\left(  {a_{1},..a_{p};b_{1},..b_{p-1};1}\right)  \propto
(1-{\sigma z)}^{-\gamma},
\end{equation}
where
\begin{equation}
-\gamma=\sum_{i=1}^{p}a_{i}-\sum_{i=1}^{p-1}b_{i}. \label{exponent}%
\end{equation}
Accordingly, high order parametrization of the hypergeometric approximants is
expected to give accurate results for the critical inverse temperature
$\beta_{c}=\frac{1}{\sigma}$ and the critical exponent $\gamma$. In table\ref{HT-SQ-xi}, we
list our results from low to high orders which shows clearly how the results
are improved in using more input information (higher orders). For instance,
our $24^{th}$ order prediction for the critical inverse temperature $\beta
_{c}=\frac{1}{\sigma}$ is $0.441068$ compared to its exact result of $0.4407$
\cite{Exactbeta}. Also, our prediction for the same order of the critical
exponent $\ \gamma=\nu\left(  2-\eta\right)  $ is $\gamma=1.76753$ compared to
its exact result $1.75$ \cite{exactexp}. Note that we considered only even orders but one can easily consider the odd ones by treating the once subtracted ($\frac{\chi-1}{\beta}$) series.

\begin{table}[H]
\centering
\caption{{\protect\scriptsize {The hypergeometric predictions for the critical inverse temperature $\beta_c$ and the critical exponent $\gamma$ with the susceptibility HT expansion (spin-half) from Ref.\cite{HT1} as input.}}}
\label{HT-SQ-xi}%
\begin{tabular}{|c|l|l|l|l|l|l|}
\hline
\multicolumn{1}{|l|}{Parameter} & \multicolumn{1}{c|}{\begin{tabular}[c]{@{}c@{}}${}_2F_1$\\  $4^{th}$ order\end{tabular}} & \multicolumn{1}{c|}{\begin{tabular}[c]{@{}c@{}}${}_3F_2$\\ $6^{th}$ order\end{tabular}} & \multicolumn{1}{c|}{\begin{tabular}[c]{@{}c@{}}${}_4F_3$\\ $8^{th}$ order\end{tabular}} & \begin{tabular}[c]{@{}l@{}}\ \ \ ${}_7F_6$\\ $14^{th}$th order\end{tabular} & \multicolumn{1}{c|}{\begin{tabular}[c]{@{}c@{}}${}_{12}F_{11}$\\ $24^{th}$ order\end{tabular}} & \multicolumn{1}{c|}{Exact} \\ \hline
$\beta_c$                       & 0.465517                                                                      & 0.450598                                                                     & 0.446082                                                                     & 0.442074                                                 & 0.441068                                                                        & 0.4407                     \\ \hline
$\gamma$                        & 1.90038                                                                       & 1.84452                                                                      & 1.82013                                                                      & 1.78474                                                  & 1.76753                                                                         & \ \ 1.75                       \\ \hline
\end{tabular}
\end{table}
 The hypergeometric series $_{p}F_{p-1}(a_{1},a_{2,}........a_{p};b_{1},b_{2},....b_{p-1};\sigma \beta)$ has a branch cut for $\beta$ along the interval $\left(1/\sigma,\infty\right)$. This means that they can probe only high temperature region ($\beta$ smaller than $1/\sigma$). The reason behind this is that the given series has coefficients with non-alternating sign. This problem is similar to Borel non-summability for a series with zero-radius of convergence. The point is that the Mellin-Barnes integral representation (Eq.(\ref{hyp-G-C2})) for the hypergeometric series will have singular points   on the contour of integration. A simple trick to overcome this problem is to use hypergeometric approximants for the series $  \chi^{-1}$ instead of approximating the $\chi$ series.  
Up to $25^{th}$ order, one can get the series expansion of $  \chi^{-1}$ as 
  
\begin{align}
 \chi^{-1}(\beta)  &  =1-4\beta+4\beta^{2}-\frac{8 }{3}\beta ^3+\dots\dots\dots-\frac{168875019032043028336 }{194896477400625} \beta
   ^{21}\nonumber\\
&  +\frac{78947836384249534384 }{43752270436875}\beta ^{22}-\frac{183378238031707095894032
  }{49308808782358125} \beta ^{23}\nonumber\\
& \frac{1151474901457395476220656 }{147926426347074375}\beta
   ^{24}
-\frac{3512146783897491495681488 }{217538862275109375}\beta ^{25}
+\dots\dots\dots \label{HTINV}. 
\end{align}  
It is now clear that the signs of the coefficients in this series are alternating and thus the corresponding  hypergeometric approximants are expected to interpolate between  high-temperature and  low-temperature regions. In Fig.1, we plot the results of three different approximations of the susceptibility in Eq.(\ref{HT}). The hypergeometric approximation (Hyp6($\beta$), left panel and Hyp24($\beta$), right panel) can probe only the high-temperature region ($\beta <\beta_c)$. The inverse hypergeometric approximation (Invhyp6($\beta$) and Invhyp24($\beta$) and the diagonal  Pad$\acute{e}$ (Pad$\acute{e}6$($\beta$) and Pad$\acute{e}24$($\beta$) ) can probe both high-temperature and low-temperature regions. In the high-temperature region, it is realized that the results of the hypergeometric and inverse hypergeometric approximations do not coincide for relatively low orders as input ( left panel for $\beta<\beta_c$). However, as we increase the order (right-panel) the results of the three approximants are almost the same. For the low-temperature region however, there is a clear difference between the inverse hypergeometric and Pad$\acute{e}$ approximations. This is expected as the Pad$\acute{e}$ approximation is not expected to give accurate results for large $\beta$ values.        
\begin{figure}[H]\label{hypinv}
\centering 
\includegraphics[width=7cm]{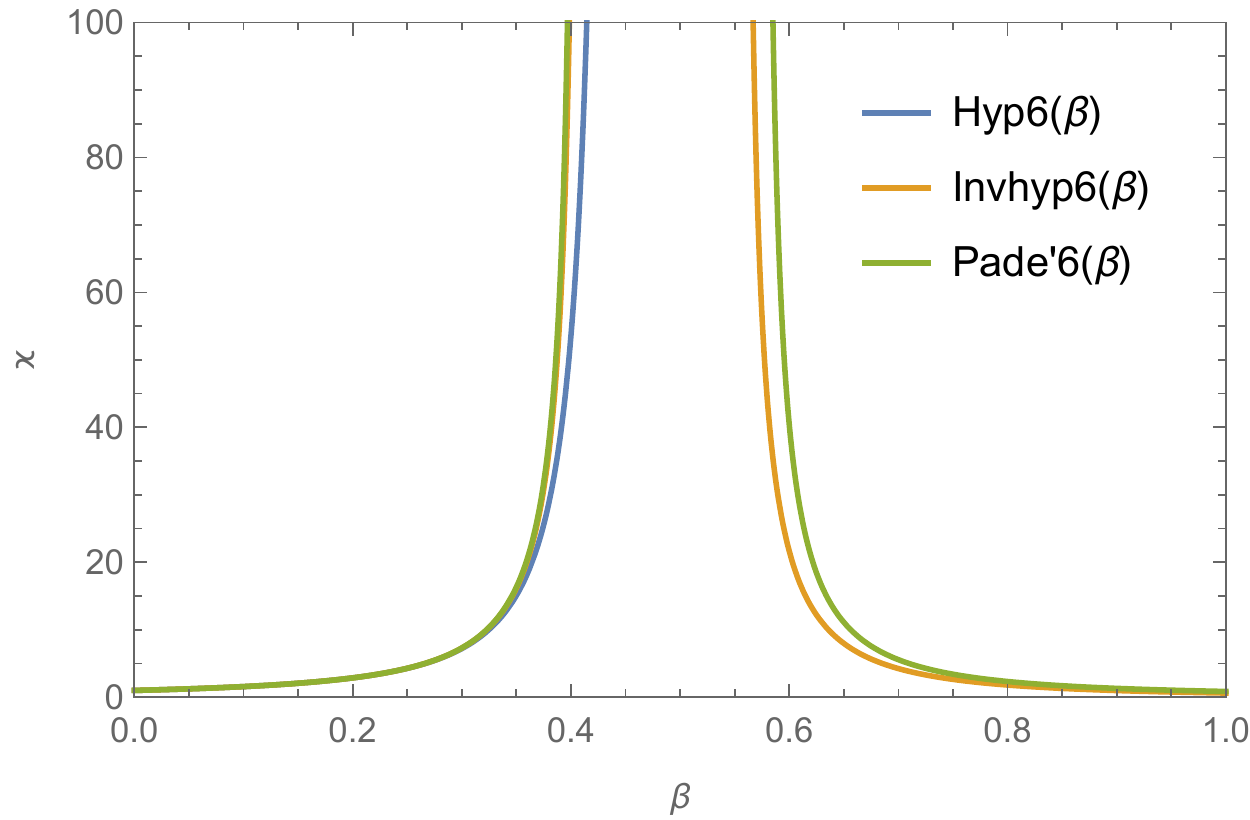}
\includegraphics[width=7cm]{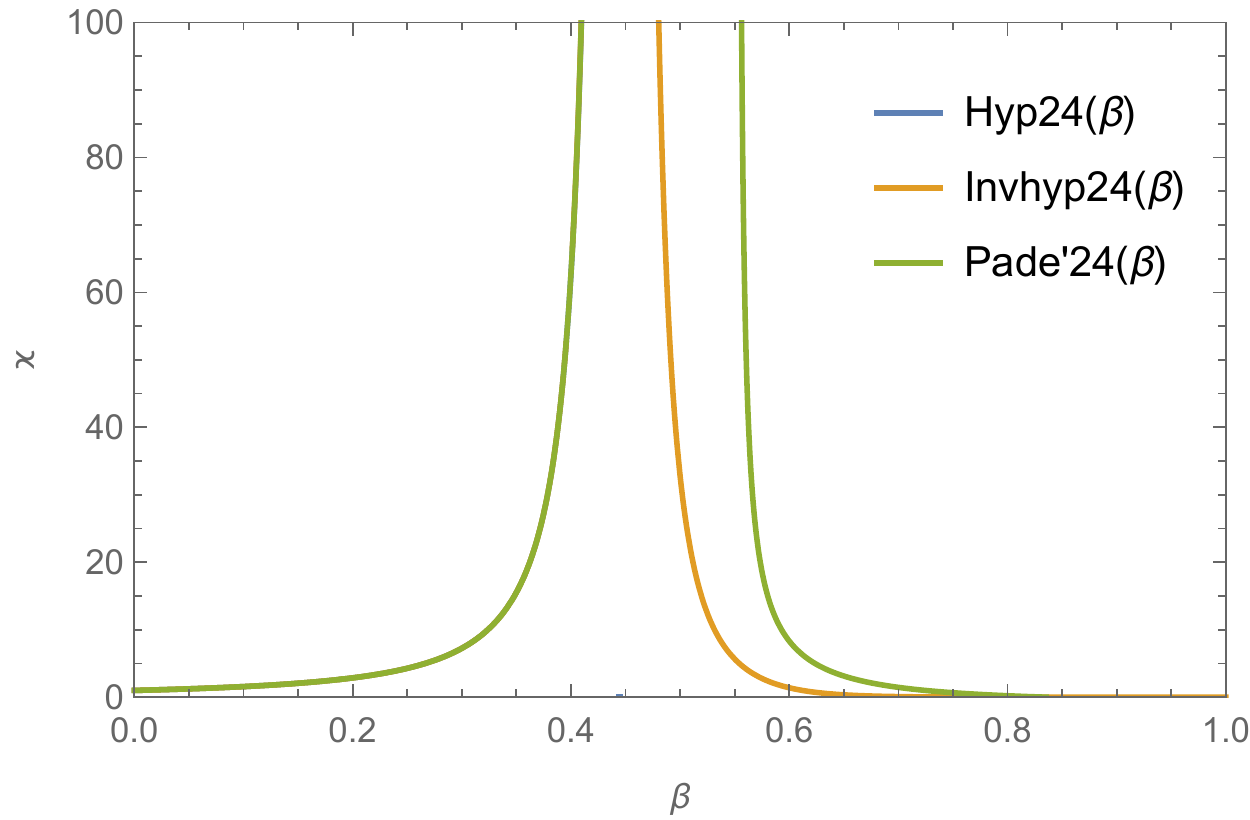} \hfill 
\caption{This figure shows the plot for three different approximations of the  six orders (left panel) and $24$ orders (right panel) series  in Eq.(\ref{HT}) for the susceptibility of the spin-half SQ-lattice. The $6^{th}$ order approximations are the hypergeometric (Hyp6($\beta$))approximation of the series in Eq.(\ref{HT}), the inverse of the hypergeometric (Invhyp6($\beta$)) approximation of Eq.(\ref{HTINV}) and the diagonal Pad$\acute{e}$ (Pad$\acute{e}6$($\beta$)) approximation of the given series in Eq.(\ref{HT}). The right panel represents the $24^{th}$ order of approximation using same three different algorithms. }  
\end{figure}
\section{Strong-Coupling and Large-order parametrization}\label{strong-large}

In the weak coupling parametrization, we used the ratio in Eq.(\ref{hnn}):
\[
g_{n}=\frac{h_{n}}{h_{n-1}}=\ \frac{%
{\displaystyle\prod_{i=1}^{p}}
\left(  a_{i}+n-1\right)  }{n%
{\displaystyle\prod_{j=1}^{q}}
\left(  b_{j}+n-1\right)  }\sigma,
\]
which has been converted to an equivalent linear set in Eq.(\ref{gnlin})
\[
\frac{h_{n}}{h_{n-1}}=\frac{\sum_{i=0}^{ p}d_{i}n^{i}}{\ \sum_{j=1}%
^{q+1}e_{j}n^{j-1}}.
\]
In many cases some non-perturbative information like large-order and
strong-coupling behaviors are available. In the field of approximating a
divergent series using techniques like Borel-Resummation, it is well known
that such non-perturbative data are able to accelerate the convergence \cite{Kleinert-Borel,Borel-6L}.
Accordingly, we need to show how to employ them in our algorithm within the
linearization technique in Eq.(\ref{gnlin}). To show this, one consider the
coefficients ratio $g_{n}$ of the hypergeometric approximant $_{p}F_{q}%
(a_{1},a_{2,}........a_{p};b_{1},b_{2},....b_{q};\sigma z)$%
\[
\frac{h_{n}}{h_{n-1}}=g_{n}=\ \frac{%
{\displaystyle\prod_{i=1}^{p}}
\left(  a_{i}+n-1\right)  }{n%
{\displaystyle\prod_{j=1}^{q}}
\left(  b_{j}+n-1\right)  }\sigma.
\]
In fact, the numerator parameters $\ \{-a_{i}\}$ are representing the parameters
of the strong-coupling expansion of the  hypergeometric series. So, for
$p>q,$ the series
\[
_{p}F_{q}(a_{1},a_{2,}........a_{p};b_{1},b_{2},....b_{q};\sigma z)=\sum
_{n=0}^{\infty}h_{n}z^{n},
\]
has the strong- coupling asymptotic form as:%

\[
_{p}F_{q}(a_{1},a_{2,}........a_{p};b_{1},b_{2},....b_{q};\sigma z)\propto
\sum_{k=1}^{p}c_{k}(-\sigma z)^{-a_{k}}%
\]

$c_{k}$ takes the form (Eq.(4.8) in Ref.\cite{Analytic2016} with $k=0$ there):%
\[
c_{k}=\frac{%
{\displaystyle\prod_{i=1,i\neq k}^{p}}
\Gamma\left(  a_{i}-a_{k}\right)  }{%
{\displaystyle\prod_{j=1\ }^{q}}
\Gamma\left(  b_{j}-a_{k}\right)  }\Gamma\left(  \ a_{k}\right).
\]
This means that if the difference between any two numerator parameters $\left(
a_{i}-a_{j}\right)  $ is an integer then $c_{k}$ is singular. Accordingly,
\ to avoid singularities in our calculations one should employ strong-coupling
parameters for which the difference $\left(  a_{i}-a_{j}\right)  $ is not an
integer. To do this, let us factorize the term $\sigma%
{\displaystyle\prod_{i=1}^{p}}
\left(  a_{i}+n-1\right)  $ into two parts:
\begin{equation}
\sigma%
{\displaystyle\prod_{i=1}^{p}}
\left(  a_{i}+n-1\right)  =\sigma%
{\displaystyle\prod_{i=1}^{l}}
\left(  a_{i}+n-1\right)  \
{\displaystyle\prod_{i=l+1}^{p}}
\left(  a_{i}+n-1\right)  , \label{factl}%
\end{equation}
assuming that for $i$ runs from $i=1$ to $i=l$, $a_{i}$ is known and the
condition $\left(  a_{i}-a_{j}\right)  \neq\mathbb{Z}$ is satisfied. Also, we assume that the large-order parameter $\sigma$ is known. If the large-order parameter $b$ is known too, one can use the relation \cite{abo-hyyp-meij,eps7,abo-precize}:%
\begin{equation}\label{largeb}
\sum_{i=1}^{p}a_{i}-\sum_{i=1}^{q}b_{i}-(p-q)=b.
\end{equation}
In the following we show how to employ the non-perturbative data for different examples.
\subsection{\texorpdfstring{$x^4$}\ \  anharmonic oscillator}
To elucidate the method of parametrization for the approximant $_{p}F_{q}(a_{1},a_{2,}........a_{p};b_{1},b_{2},....b_{p-2};\sigma z)$ using weak-coupling, strong-coupling and large-order data,
let us first take an   example of considering the parametrization of a
the series  up to the $6^{th}$ order which then approximated
by $_{6}F_{4}(a_{1},a_{2,}\ ......a_{6,};b_{1},b_{2},....b_{4};\sigma z)).$ For  the anharmonic oscillator series in Eq.(\ref{x4-pert}), the strong-coupling expansion goes like \cite{harm-sigma}
\[
E_{0}(g)=c_{1}g^{\frac{1}{3}}+c_{2}g^{-\frac{1}{3}}+c_{3}g^{-1}+c_{4}%
g^{-\frac{5}{3}}+.....
\]
To avoid singularities, one can employ $a_{1}=\frac{-1}{3},a_{2}=\frac{1}{3}$
and $a_{3}\ =1$ or in other words $l$ in Eq.(\ref{factl}) is $3$. Also, from
Ref.\cite{harm-sigma}, $\sigma=-3$ and $b=\frac{-1}{2}$. Accordingly,
Eq.(\ref{factl}) takes the from
\begin{equation}
\sigma%
{\displaystyle\prod_{i=1}^{p}}
\left(  a_{i}+n-1\right)  =\left(-3\ \left(  \frac{-1}{3}+n-1\right)  \left(
\frac{1}{3}+n-1\right)  \left(  1+n-1\right) \right) \
{\displaystyle\prod_{i=4}^{6}}
\left(  a_{i}+n-1\right)  .
\end{equation}
As we did before in weak-coupling parametrization, the term $%
{\displaystyle\prod_{i=4}^{6}}
\left(  a_{i}+n-1\right)  \ $ can be rewritten in a linear form as:%
\[%
{\displaystyle\prod_{i=4}^{6}}
\left(  a_{i}+n-1\right)  =d_{3}n^{3}+d_{2}n^{2}+d_{1}n+d_{0},
\] where $d_3=1$.
Also, we have
\[
n%
{\displaystyle\prod_{j=1}^{4}}
\left(  b_{j}+n-1\right)  =\left(  n^{5}+\sum_{j=1}^{4}e_{j}n^{j}\right),
\]
with  $d_{2}=\left(  a_{4}+a_{5}+a_{6}-3\right)  .$ We can use the
formula:
\begin{equation}
\sum_{i=1}^{p}a_{i}-\sum_{i=1}^{p-2}b_{i}-2=b,
\end{equation}
to express $e_{4}$ in terms of $a_{i}$ and $b$. Or
\[
\frac{-1}{3}+\frac{1}{3}+1+d_{2}+3-2+\frac{1}{2}=e_{4}+4,
\]

or
\[
e_{4}=d_{2}-\frac{3}{2}.
\]
One can generalize this formula \ to any $l$ and $b$ as
\begin{equation}
\sum_{i=1}^{l}a_{i}+d_{p-l}-(b+l)=e_{p-2}\ . \label{d-e-rel}%
\end{equation}
Accordingly, for the parametrization of the hypergeometric approximant 
	\[
_{6}F_{4}(a_{1},a_{2,}\ ......a_{6,};b_{1},b_{2},....b_{4};\sigma z)),
\]
 we only need six orders from the weak-coupling series ( the approximant has $11$
parameters).   The parametrization then will go through  the set of  equations:
\begin{align*}
&  \left(  -3\ \right)  \left(  \frac{-1}{3}+n-1\right)  \left(  \frac{1}%
{3}+n-1\right)  \left(  \frac{5}{3}+n-1\right)  \left(  \ n^{3}+d_{2}%
n^{2}+d_{1}n+d_{0}\right) \\
&  =R_{n}\left(  n^{5}+\left(  d_{2}-\frac{5}{6}\right)  n^{4}+e_{3}%
n^{3}+e_{2}n^{2}+e_{1}n\right)  ,
\end{align*}
which is a set of linear equations in the six unknowns $d_{2},d_{1},d_{0}%
,e_{3},e_{2},e_{1}$. Note that from Eq.(\ref{x4-pert}) we have
\[
R_{1}=\frac{\frac{3}{4}\ }{\frac{1}{2}}=\frac{3}{2},\ R_{2}=\frac{\frac
{-21}{8}\ }{\frac{3}{4}}=-\frac{7}{2},........
\]
The unknown numerator and denominator parameters are thus obtained from the roots
of the polynomials:%
\begin{align*}
\left(  1-A\right)  ^{3}+\sum_{i=0}^{2}d_{i}\left(  1-A\right)  ^{i}  &  =0,\\
e_{1}\ +e_{2}\left(  1-B\right)  \ +e_{3}\left(  1-B\right)  ^{2}+\left(
d_{2}-\frac{3}{2}\right)  \left(  1-B\right)  ^{3}+\left(  1-B\right)  ^{4}
&  =0.
\end{align*}
With the non-perturbative information $a_{1}=\frac{-1}{3},a_{2}=\frac{1}%
{3},a_{3}\ =1,\sigma=-3$ and $b=\frac{-1}{2}$ for any $n$ orders of the given
series are known, these equations can be generalized as
$\ $%
\begin{align}
\left(  1-A\right)  ^{p-3}+\sum_{i=0}^{p-4}d_{i}\left(  1-A\right)  ^{i} &
=0,\nonumber\\
\left(  1-B\right)  ^{p-2}+\left(  d_{p-4}-\frac{3}{2}\right)  \left(
1-B\right)  ^{p-3}+\sum_{j=1}^{p-3}e_{j}\left(  1-B\right)  ^{j-1} &
=0,\label{set-lin-S-b}%
\end{align}
with $2p-6=n$. This equation can be generalized to any   $n$ perturbative
terms and $l$ strong-coupling parameters    of the input series where then it takes the form:
\begin{align}
\left(  1-A\right)  ^{p-l}+\sum_{i=0}^{p-l-1}d_{i}\left(  1-A\right)  ^{i} &
=0,\nonumber\\
\left(  1-B\right)  ^{p-2}+e_{p-2}\left(  1-B\right)  ^{p-3}+\sum_{j=1}%
^{p-3}e_{j}\left(  1-B\right)  ^{j-1} &  =0,
\end{align}
where  $p=\frac{1}{2}n+\frac{1}{2}l+\frac{3}{2}\ $and
\begin{equation}
\sum_{i=1}^{l}a_{i}+d_{p-l}-(b+l)=e_{p-2}.
\end{equation}
One can double check the algorithm by solving directly (for time considerations take low orders) the non-linear set of
the form:%
\begin{align*}
R_{n} &  =\frac{\left(  -3\ \right)  \left(  \frac{-1}{3}+n-1\right)  \left(
\frac{1}{3}+n-1\right)  \left(  \frac{5}{3}+n-1\right)
{\displaystyle\prod_{i=4}^{p}}
\left(  a_{i}+n-1\right)  }{n%
{\displaystyle\prod_{j=1}^{q}}
\left(  b_{j}+n-1\right)  },\\
b &  =\sum_{i=1}^{p}a_{i}-\sum_{i=1}^{q}b_{i}-2,\
\end{align*}
and compare the results. In fact, we did that and found exactly the same results. In table \ref{x4S}, we listed our prediction for the ground state energy and it can be seen from the table how fast one can get accurate results from a relatively low order in employing the non-perturbative data in the parametrization process. For instance, the result of 14th order parametrization of $E_0(g)$    at  $g=50$ shares the first five digits with the exact result. This accuracy has been obtained at the 25th order of the weak-coupling parametrization in table \ref{x4w}.

\begin{table}[H]
\centering 
\caption{{\protect\scriptsize { The predictions  for the ground state energy of the  $x^4$ anharmonic oscillator using  our hypergeometric approximants (fed  with non-perturbative parameters $a_1=-\frac{1}{3},a_2=\frac{1}{3},a_3=1,b=-\frac{1}{2},\sigma=-3$) compared to  numerical results $E_{exact}$ from Ref.\cite{har4}.}}}%
\label{x4S}%
\begin{tabular}{|c|c|c|c|c|c|}
\hline
g   & \begin{tabular}[c]{@{}c@{}}${}_5F_3$\\  4th order\end{tabular} & \begin{tabular}[c]{@{}c@{}}${}_6F_4$\\ 6th order\end{tabular} & \begin{tabular}[c]{@{}c@{}} ${}_8F_8$\\ 10th order\end{tabular} & \begin{tabular}[c]{@{}c@{}} ${}_{10}F_8$\\ 14th order\end{tabular} & Exact         \\ \hline
0.1 & 0.5591497782                                             & 0.559146314                                             & 0.5591463275                                             & 0.5591463272                                              & 0.5591463272  \\ \hline
0.5 & 0.6963160313                                             & 0.6961755535                                            & 0.6961760318                                             & 0.6961758243                                              & 0.696 1758208 \\ \hline
1   & 0.8041767370                                             & 0.8037711222                                            & 0.8037718820                                             & 0.8037706822                                              & 0.8037706512  \\ \hline
2   & 0.9525031038                                             & 0.9515730530                                            & 0.9515731615                                             & 0.9515686334                                              & 0.9515684727  \\ \hline
50  & 2.508468788                                              & 2.499878593                                             & 2.499831098                                              & 2.499717261                                               & 2.4997087726  \\ \hline
\end{tabular}
\end{table}
 
\subsection{The \texorpdfstring{$\mathcal{PT}-$}\ symmetric \texorpdfstring{$ix^{3}$}\ \  anharmonic oscillator}
For the series in Eq.(\ref{pertub}), the associated large-order asymptotic
behavior is given in Ref. \cite{Zinnx3} as :
\begin{equation}
c_{n}\sim\alpha n!(-\sigma)^{n}n^{b}\left(  1+O\left(  \frac{1}{n}\right)
\right)  ,
\end{equation}
where $b=-\frac{1}{2},\sigma=-\frac{5}{24}$. Also, the strong-coupling
behavior of the ground state energy is  given in the same reference where the first $5$   parameters are given as :%
\[
a_{1}=1,a_{2}=-\frac{1}{5},a_{3}=\frac{3}{5},a_{4}=\frac{7}{5},a_{5}%
=\frac{11}{5}.
\]
Accordingly, the $(2p-8)^{th}$ order of the series in Eq.(Eq.(\ref{pertub})) can be
parametrized using the hypergeometric approximant $_{p}F_{p-2}(a_{1}%
,a_{2,}........a_{p};b_{1},b_{2},....b_{p-2};\sigma z)$ where
Eq.(\ref{d-e-rel}) now reads:
 \begin{equation}
d_{p-6}+\frac{1}{2}=e_{p-2}\ .
\end{equation}
 We list the results in table \ref{x3S} where one can see that they are competitive to exact ones at the $14th$ order and it is more accurate than  the $19th$ order parametrization of  the weak coupling parametrization in table \ref{x3w}. 
\begin{table}[H]
\centering
\caption{{\protect\scriptsize {  The results for the weak-coupling, strong coupling and large-order parametrization of the  hypergeometric approximants for the ground state energy of the $\mathcal{PT}$-symmetric $ix^3$ theory. The results are compared to numerical calculations from Ref.\cite{ix3-bender-large} (with $g\equiv 288\lambda^2$).}}}%
\label{x3S}%
\begin{tabular}{|l|l|l|l|l|l|l|}
\hline
\multicolumn{1}{|c|}{g}       & \multicolumn{1}{c|}{\begin{tabular}[c]{@{}c@{}}${}_7F_5$\\  6th order\end{tabular}} & \multicolumn{1}{c|}{\begin{tabular}[c]{@{}c@{}}${}_9F_7$\\ 10th order\end{tabular}} & \multicolumn{1}{c|}{\begin{tabular}[c]{@{}c@{}}${}_{10}F_8$\\ 12th order\end{tabular}} & \multicolumn{1}{c|}{\begin{tabular}[c]{@{}c@{}}${}_{11}F_9$\\ 14th order\end{tabular}} & \begin{tabular}[c]{@{}l@{}}${}_{12}F_{10}$\\ 16th order\end{tabular} & \multicolumn{1}{c|}{Exact}   \\ \hline
\multicolumn{1}{|c|}{0.28125} & \multicolumn{1}{c|}{0.509741}                                                 & \multicolumn{1}{c|}{0.509976}                                                 & \multicolumn{1}{c|}{0.510165}                                                  & \multicolumn{1}{c|}{0.509976}                                                  & 0.509975                                                   & \multicolumn{1}{c|}{0.50998} \\ \hline
\multicolumn{1}{|c|}{1. 125}  & \multicolumn{1}{c|}{0.536976}                                                 & \multicolumn{1}{c|}{0.536976}                                                 & \multicolumn{1}{c|}{0.536976}                                                  & \multicolumn{1}{c|}{0.536976}                                                  & 0.536976                                                   & \multicolumn{1}{c|}{0.53393} \\ \hline
\multicolumn{1}{|c|}{4.5}     & \multicolumn{1}{c|}{0.594923}                                                 & \multicolumn{1}{c|}{0.594915}                                                 & \multicolumn{1}{c|}{0.594915}                                                  & \multicolumn{1}{c|}{0.594915}                                                  & 0.594915                                                   & \multicolumn{1}{c|}{0.59492} \\ \hline
\ \ \ 18                            & 0.713073                                                                      & 0.712940                                                                      & 0.712935                                                                       & 0.712936                                                                       & 0.712936                                                   & 0.71294                      \\ \hline
\ \ \ 72                            & 0.901015                                                                      & 0.900293                                                                      & 0.900254                                                                       & 0.900259                                                                       & 0.900258                                                   & 0.90026                      \\ \hline
\ \ \ 288                           & 1.16958                                                                       & 1.16758                                                                       & 1.167440                                                                       & 1.167458                                                                       & 1.167455                                                   & 1.16746                      \\ \hline
\ \ \ 1152                          & 1.53490                                                                       & 1.53105                                                                       & 1.53080                                                                        & 1.53077                                                                        & 1.53077                                                    & 1.53078                      \\ \hline
\end{tabular}
\end{table}
\section{weak-coupling, Strong-coupling and large order parametrization of a series with finite radius of convergence}\label{strong-large-f}
The above recipe can also be modified to parametrize   the approximant
	\[
	 _{p}F_{p-1}%
(a_{1},a_{2,}........a_{p};b_{1},b_{2},....b_{p-1};\sigma z), 
\]
for the analytic continuation of a series with finite radius of convergence. Let us  for instance consider the series in
Eq.(\ref{Jpert}) which  has $b=\frac{-3}{2}$ \cite{Large-strong} while the strong $J$
behavior has the parameters (with non-integer difference)\cite{Zinnx3}:
\[
a_{1}=-\frac{3}{2},\ a_{2}=-\frac{1}{4},\ a_{3}=1,\ \ a_{4}=\frac{9}{4}\ .
\]
The suitable approximant for that series is $_{p}F_{p-1}(a_{1},a_{2,}%
........a_{p};b_{1},b_{2},....b_{p-1};\sigma z)$, where we solve the set :%
\begin{align}
\frac{%
{\displaystyle\prod_{i=1}^{p}}
\left(  a_{i}+n-1\right)  }{n%
{\displaystyle\prod_{j=1}^{q}}
\left(  b_{j}+n-1\right)  }\sigma & =\frac{c_{n}}{c_{n-1}}\text{ or
equivelantely }\nonumber\\
\ n\left(  n-\frac{5}{2}\right)  \left(  n-\frac{5}{4}\right)  \left(
n+\frac{5}{4}\right)  \sum_{i=0}^{p-4}d_{i}n^{i}  & =\frac{c_{n}}{c_{n-1}}\sum_{j=1}^{p}e_{j}%
n^{j},
\end{align}
with $e_{p}=1$ , $e_{p-1}=d_{p-5}-1$ and $d_{p-4}=\sigma$ ($\sigma$ is not known exactly  for this model). To get the
unknown parameters we solve the following polynomial equations:%

\begin{align}
\ \sum_{i=0}^{p-4\ }d_{i}\left(  1-A\right)  ^{i} &  =0\text{, with }%
a_{i}\text{ are the roots,}\nonumber\\
\left(  1-B\right)  ^{p-1}+e_{p-1}\left(  1-B\right)  ^{p-2}+\sum_{j=1}%
^{p}e_{j}\left(  1-B\right)  ^{j-1} &  =0\text{, with }b_{i}\text{ are the
roots,}%
\end{align}
If we have $n$ orders of the perturbation series, then the value of $p$ is determined from the relation $p=$ $\frac
{1}{2}n+\frac{5}{2}$.
The $19th$ order has been parametrized using the non-perturbative data and compared to   orders without  non-perturbative information in table \ref{x3-lee-w-Et}. When compared with the $150th$ order of continued fraction method in Ref. \cite{Zinnx3}, the non-perturbative parametrization are more accurate than the weak-coupling only parametrization. 

\begin{table}[H]
\centering
\caption{{\protect\scriptsize {The results of the weak-coupling, strong-coupling and large-order parametrization ($ 4^{th}$ column) of the hypergeometric approximant   for the ground state energy for the model in Eq.(\ref{Hamilt}). The results are  compared to those from Continued Fraction method ($150$th order in Ref.\cite{Zinnx3}) and weak-coupling-only parametrization.}}}
\label{x3-lee-w-Et}%
\begin{tabular}{|c|llll|}
\hline
\multicolumn{1}{|l|}{}  & \multicolumn{4}{c|}{$E_0^J$}                                                                                                                                                                                                                                                             \\ \hline
\multicolumn{1}{|l|}{\ \ \ \ J} & \multicolumn{1}{c|}{\begin{tabular}[c]{@{}c@{}}${}_5F_4$ \\ 10th order\end{tabular}} & \multicolumn{1}{c|}{\begin{tabular}[c]{@{}c@{}}${}_{10}F_9$ \\ 20th order\end{tabular}} & \multicolumn{1}{c|}{\begin{tabular}[c]{@{}c@{}}${}_{12}F_{11}$ \\ 19th order\\ with non-perturbative\\ parameters\end{tabular}} & \multicolumn{1}{c|}{Continued Fraction} \\ \hline
$-2^{4/5}$              & \multicolumn{1}{l|}{0.395189 -0.298946i}                                      & \multicolumn{1}{l|}{0.389793-0.363668i}                                       & \multicolumn{1}{l|}{0.389823 - 0.364412 i}                                       & 0.3898(5)-0.3644(3) i                   \\ \hline
$-5^{-4/5}$                                                    & \multicolumn{1}{l|}{0.282699258193271}                                        & \multicolumn{1}{l|}{0.2826992581963768}  & \multicolumn{1}{l|}{0.28269925819333014}                                       & 0.2826992581932749098990(1)             \\ \hline
-1                                                             & \multicolumn{1}{l|}{0.1957508231732275}                                       & \multicolumn{1}{l|}{0.19575081572380823}& \multicolumn{1}{l|}{0.19575081571720868}                                       & 0.195750 8157161719(6)                  \\ \hline
$-21.6^{-4/5}$                                                 & \multicolumn{1}{l|}{0.3421580186193393}                                       & \multicolumn{1}{l|}{0.3421580186203626} & \multicolumn{1}{l|}{0.3421580186193456}                                       & 0.34215801861934042140767(6)            \\ \hline
\end{tabular}
\end{table}
\section{Summary and Conclusions}\label{conc}
The hypergeometric-Meijer approximation algorithm has been introduced and applied in our previous work \cite{eps7,universal,abo-precize,abo-hyyp-meij}. The algorithm proved to be accurate and one can realize, for instance, that the specific heat exponent extracted from approximating the renormalization group series of the $O(2)$ scalar field model \cite{eps7,lambda} has been reached an accuracy that has never been met before (for the same series). Besides, with the aid of the Mathematica Built-in Meijer-G functions, the algorithm  proved to be the simplest approximation algorithm when compared to more sophisticated ones that can compete its accuracy. The algorithm is also able to approximate series with different analytic properties.  Moreover, the algorithm has been shown to accommodate every possible information about the given series like weak-coupling, strong-coupling and large-order data.

The major problem  with the previous version of the algorithm is that one has to solve a set of non-linear equations and the degree of non-linearity increases with increasing the order of the input series. Practically, it will be very hard and might be impossible to use normal PC to solve the set for orders higher than seven. In this work, we were able to build up an equivalent (order by order) linear set of equations that makes the calculations go very fast. We also are able to employ the non-perturbative data to accelerate the convergence of the approximation algorithm within the new version.

 We stressed  different type of applications to show that the current version of the algorithm is fast, accurate and simple as well. The type of applications stressed in this work are divided into two main parts: i) parametrization  of the approximants using only the weak-coupling data as input ii)   parametrization that uses all weak-coupling, strong-coupling and large-order data. For the first part, we applied the algorithm to a series with finite-radius of convergence as well as a series with zero-radius of convergence. The corresponding  approximants were able to predict very accurate results for the non-perturbative parameters using only perturbative information. For the other part of the applications,   the non-perturbative data were able to accelerate the convergence of the approximation algorithm.    

For the weak-coupling parametrization, we considered  two examples for a series with-zero radius of convergence. In the first example we considered the approximation of the series for the ground state energy of the $x^4$ anharmonic oscillator. The approximation (specially for high orders) gives accurate results. We have shown in previous work  \cite{abohyp,eps7,universal,abo-precize,abo-hyyp-meij} that one can extract the non-perturbative data from the weak-coupling parametrization of the hypergeometric approximants. The predictions in table \ref{x4wp} shows a very accurate results for the strong-coupling parameter $S^*$ and the large-order parameters $\sigma$ and $b$. However, accurate results for $b$ can be attained only for relatively high orders as input. Although for some orders the approximants are singular which means that it leads to no prediction for the ground state energy for that order, the prediction of the non-perturbative parameters are accurate. The point is that while for high orders the hypergeometric approximant is very accurate in representing the given series, the analytic continuation using the Mellin-Barnes integral in Eq.(\ref{hyp-G-C2}) might not work. In other words, the conditions needed for a finite integral may not be satisfied. 

The other example for a series with zero-radius of convergence studies the approximation of the ground state energy of the $\mathcal{PT}$-symmetric $ix^3$ model. Again accurate results have been obtained for the ground state energy   (table \ref{x3w})  and  for the non-perturbative parameters as well.    
     
 The hypergeometric algorithm  have been used also for the analytic continuation of a   series with finite-radius of convergence outside the convergence   disc. We applied the algorithm  for the approximation of two series of that type. First, we considered the ground state energy of the Yang-Lee model (strong-coupling expansion of $ix^3$ model). The approximants astonishingly give very accurate results either for the ground  state energy or the non-perturbative parameters. Note that all of these predictions used only weak-coupling information as input. Second, we stressed  the high-temperature expansion for the susceptibility of the spin-half Ising model (SQ Lattice). Accurate predictions for the critical temperature and critical exponent have been also obtained. 

The hypergeometric approximant parametrized by high-temperature data fails to describe the low-temperature behavior. The point is that all coefficients of the high-temperature series expansion are positive and a problem like non-Borel summability has been faced. To overcome this problem, we extracted the expansion of $\chi^{-1}$ from that of $\chi$ hopping to obtain a series with coefficients  alternating in sign. The idea worked out and the hypergeometric approximants for the new series is able to probe the low-temperature region although it has fed with only high-temperature information. Moreover, for relatively high order the approximants of $\chi$ series and $\chi^{-1}$ series almost coincide for the high temperature region but the approximants for $\chi^{-1}$ extrapolates to the low-temperature region. 

Except of the high temperature expansion, all the examples studied in this work for the weak coupling case have known non-perturbative data. We adapted the algorithm to accommodate such parameters to accelerate the convergence of the approximation algorithm. The corresponding predictions (listed in Sec. \ref{strong-large}) fit with  that expectation  very clearly.

We can claim that the version of the hypergeometric-Meijer approximation algorithm introduced in this work is fast, simple and accurate to the extent that might make it one of  the most preferred   approximation algorithms. It can be said also that it is an all-in-one algorithm in the sense that it can approximate different type of series with different growth factors ($0!,n!, (2n)!,....$. Moreover, it can accommodate all kinds of available information for the given series either perturbative or non-perturbative ones. Also, one of its astonishing features is the extraction  of accurate predictions for the non-perturbative parameters  of the given series with only perturbative data as input
\bibliography{Parametrization}
 \end{document}